\documentclass[12pt,preprint]{emulateapj}

\usepackage{graphicx,amssymb,amsmath,times}
\usepackage{longtable}
\usepackage{amsmath, amssymb} 
\usepackage{url}
\usepackage[varg]{txfonts}
\usepackage[T1]{fontenc}
\usepackage[latin9]{inputenc}
\usepackage{multirow}
\usepackage{natbib}
\usepackage{txfonts}

\def\simlt{\lower.5ex\hbox{\ltsima}}
\def\simgt{\lower.5ex\hbox{\gtsima}}

\def\farcm{\hbox{$\mkern-4mu^\prime$}}

\def\farcs{\hbox{$^{\prime\prime}$}~}

\def\gtsim{\;\lower.6ex\hbox{$\sim$}\kern-6.7pt\raise.4ex\hbox{$>$}\;}
\def\ltsim{\;\lower.6ex\hbox{$\sim$}\kern-6.9pt\raise.4ex\hbox{$<$}\;}
\def\hr{${}^{\hbox{\footnotesize h}}$}
\def\tm{${}^{\hbox{\footnotesize m}}$}
\def\ts{${}^{\hbox{\footnotesize s}}$}
\def\Ts{${}^{\hbox{\footnotesize s}}$\llap{.}}

\def\Sec{${}^{\prime\prime}$\llap{.}}
\def\deg{${}^\circ$}
\def\min{${}^{\prime}$}
\def\sec{${}^{\prime\prime}$}

\def\bmv{\hbox{\it B--V\/}}
\def\bmr{\hbox{\it B--R\/}}
\def\bmi{\hbox{\it B--I\/}}

\def\jmk{\hbox{\it J--K\/}}

\def\ngc#1{\hbox{NGC$\,$#1}}

\shorttitle{Photometry for RR~Lyraes in M4}
\shortauthors{Stetson et~al.}

\begin{document}

\title{Optical and Near-Infrared {\it UBVRIJHK\/} Photometry for
the RR~Lyrae stars in the Nearby Globular Cluster M4
(NGC~6121)\textsuperscript{*}}

\footnotetext[*]{Based in part on data obtained from the ESO
Science Archive Facility under muliple requests by the authors; in
part on data obtained from the Isaac Newton Group Archive, which is
maintained as part of the CASU Astronomical Data Centre at the
Institute of Astronomy, Cambridge; and in part upon data
distributed by the NOAO Science Archive. NOAO is operated by the
Association of Universities for Research in Astronomy (AURA) under
cooperative agreement with the National Science Foundation.  This
research also benefited from the Digitized Sky Survey service
provided by the Canadian Astronomy Data Centre operated by the
National Research Council of Canada with the support of the
Canadian Space Agency.\newline}

\author{P.~B.~Stetson\altaffilmark{1}, 
V.~F.~Braga\altaffilmark{2}, 
M.~Dall'Ora\altaffilmark{3}, 
G.~Bono\altaffilmark{2,4}, 
R.~Buonanno\altaffilmark{2,5}, 
I.~Ferraro\altaffilmark{4}, 
G.~Iannicola\altaffilmark{4}, 
M.~Marengo\altaffilmark{6}, 
J.~Neeley\altaffilmark{6}}

\altaffiltext{1}{NRC-Herzberg, Dominion Astrophysical Observatory, 5071 West Saanich Road, Victoria BC V9E 2E7, Canada}
\altaffiltext{2}{Department of Physics, Universit\`a di Roma Tor Vergata, via della Ricerca Scientifica 1, 00133 Roma, Italy}
\altaffiltext{3}{INAF-Osservatorio Astronomico di Capodimonte, Salita Moiariello 16, 80131 Napoli, Italy}
\altaffiltext{4}{INAF-Osservatorio Astronomico di Roma, via Frascati 33, 00040 Monte Porzio Catone, Italy}
\altaffiltext{5}{INAF-Osservatorio Astronomico di Teramo, Via Mentore Maggini snc, Loc. Collurania, 64100 Teramo, Italy}
\altaffiltext{6}{Department of Physics and Astronomy, Iowa State University, Ames, IA 50011, USA}

\date{\centering Submitted \today\ / Received / Accepted }

\begin{abstract}

We present optical and near-infrared {\it UBVRIJHK\/} photometry
of stars in the Galactic globular cluster M4 (\ngc{6121}) based
upon a large corpus of observations obtained mainly from public
astronomical archives.  We concentrate on the RR~Lyrae variable
stars in the cluster, and make a particular effort to accurately
reidentify the previously discovered variables.  We have also
discovered two new probable RR~Lyrae variables in the M4 field:
one of them by its position on the sky and its photometric
properties is a probable member of the cluster, and the second
is a probable background (bulge?) object.  We provide accurate
equatorial coordinates for all 47 stars identified as RR Lyraes,
new photometric measurements for 46 of them, and new period estimates
for 45.  We have also derived accurate positions and mean
photometry for 34 more stars previously identified as variable
stars of other types, and for an additional five non-RR Lyrae
variable stars identified for the first time here.  We present
optical and near-infrared color-magnitude diagrams for the
cluster and show the locations of the variable stars in them.  We
present the Bailey (period-amplitude) diagrams and the
period-frequency histogram for the RR Lyrae stars in M4 and
compare them to the corresponding diagrams for M5 (\ngc{5904}).  We conclude
that the RR~Lyrae populations in the two clusters are quite
similar in all the relevant properties that we have considered. 
The mean periods, pulsation-mode ratios, and Bailey diagrams of
these two clusters show support for the recently proposed
``Oosterhoff-neutral'' classification.  

\end{abstract}
\keywords{Stars; Star Clusters and Associationss}

\maketitle

\section{Introduction} \label{chapt_intro}

The Galactic globular clusters (GGCs) \ngc{6397} and  M4
(\ngc{6121}) play a key role among all the GGCs, since these are
the closest two globulars to the Sun. Unlike \ngc{6397}, M4 is
located near the equator ($\alpha\,$=$\,$16\hr$\,$23\tm$\,$35\ts,
$\delta\,$=$\,$--26\deg$\,$31\min$\,$32\sec), making it a good
target for both northern and southern observing facilities.  Again
unlike \ngc{6397}, M4 hosts a significant population of RR~Lyrae
variables, which offers a whole new class of observational
constraints on the cluster's distance and physical properties. 
The main drawback of M4 is that it is veiled by large and
differential reddening: E(\bmv)$\,=\,$0.37$\pm$0.01 (estimated
standard error of the mean, overall range $\delta$E(\bmv)$\gtsim$
0.2\,mag) according to \cite{hendricks2012} (hereinafter H12); see
also \cite{ivans1999}, \cite{marino2008} and
\cite{mucciarelli2011}.  Furthermore, several studies in recent
years have suggested that the reddening law toward M4 is abnormal
compared to the canonical diffuse interstellar medium
\citep{cardelli,mccall2004}.   These drawbacks are due to the fact
that M4 is located behind the Scorpius-Ophiuchus cloud complex.
The problem of the reddening law in the direction of M4 has been
recently discussed in a thorough investigation by H12 on the basis
of both optical and near-infrared (NIR) photometry.  

Its small distance makes M4 a valuable laboratory for stellar
evolution studies, since even modest telescopes can provide accurate
photometric and spectroscopic data well below the main-sequence turnoff.
The same consideration applies to variable stars, and indeed M4
hosts a sizable sample of stars identified as fundamental (FU, 31) and first overtone
(FO, 13) RR~Lyraes \citep{clement2001}, and optical photoelectric
photometry has been published for a significant fraction of them
\citep{sturch1977,cacc1979,liujanes1990,clementini1994}.  A
sizable sample of eclipsing binaries has been detected and studied by
\cite{kaluz1997}, \cite{moche2002}, \cite{kaluz2013a} and 
\cite{kaluz2013b}.  A detailed X-ray and optical investigation of
the innermost cluster regions was performed by
\cite{bassa2004}, who detected a dozen chromospherically
active binaries, confirmed the presence of a millisecond pulsar
\citep{lyne1988,sigurd2003}, and identified a couple of candidate
cataclysmic variables.  More recently, deep optical 
imaging with the Advanced Camera for Surveys (ACS) on the 
{\it Hubble Space Telescope\/} (HST) has provided a very deep
color-magnitude diagram approaching the hydrogen~burning limit on the main
sequence, and showing the blue hook on the white dwarf cooling 
sequence \citep{hansen2004}.    

A deep and accurate optical--NIR CMD based on images collected with HST 
was provided by \citet{milone2014}. They found evidence of a split in 
the main sequence region located below the knee. In particular, they found 
that one subpopulation can be associated 
with normal chemical abundance ratios, and the other with a composition that 
is enhanced in nitrogen and depleted in oxygen.     

This result appears even more compelling in conjunction with the
results of a very detailed spectroscopic analysis of evolved and
main-sequence stars performed by \citet{malavolta2014}. They
collected a large number of high-resolution spectra for more than
two thousand cluster members and found that the overall
metallicity is almost constant over the entire sample:
[Fe/H]$\,=\,$--1.07, $\sigma\,$=$\,0.02$. Moreover, the difference in iron
content between the two distinct stellar sequences along the RGB of
M4, identified by \citet{monelli2013}, is smaller than 0.01 dex. 
This is consistent with the evidence from \citet{milone2014} that the
difference appears to be in the CNO abundances.  

A deep and accurate NIR CMD of M4 was also provided by
\citet{libralato2014}, who analysed NIR images collected with
HAWK-I at VLT.  Their data cover a time interval of eight years,
and accurate ground-based astrometry allowed them to distinguish
cluster and field stars.  

This is a paper in a series dedicated to multiband photometry of
cluster and field RR~Lyrae stars.  In \S\S 2 and 3 below, we
present both optical and NIR data sets for the variable stars of
M4.  Then \S4 deals with the reidentification and characterization of
the variable stars in the cluster field, while \S5 considers them in the
context of the cluster color-magnitude diagram.  After that, \S6 discusses the
pulsation properties of the RR~Lyrae variables and the implications for their
evolutionary status, and \S7 compares and contrasts the RR~Lyrae
population of M4 to that of another well-studied cluster, M5 (\ngc{5904}).
Finally, \S8 briefly summarizes the results of our study.  

\section{Optical photometry}\label{chapt_obs_opt}

Our optical observations of M4 consist of 5,003 individual CCD
images obtained during the course of 18 observing runs. 
Table~\ref{tab:table_obs_opt} is a synopsis of the observation
dates and the number of exposures obtained in each filter during
each run.  The column ``multiplex'' refers to the number of
individual CCD chips in the particular camera, which were treated
as independent detectors.  The multiplex factor times the number
of exposures represents the net contribution to the grand total of
5,003 CCD images cited above.  

These observations were included in a total of 84 datasets,
where a dataset is defined as either (a)~all the digital images
obtained with a single CCD chip on a single photometric night, or
(b)~all the digital images obtained with a single CCD on one or
more non-photometric nights during the same observing run.
Datasets from photometric nights are calibrated with respect to
standard stars in many fields spread across the sky, including
full corrections for atmospheric extinction.  Datasets from
non-photometric nights are calibrated with respect to local
standards contained within each image, which have themselves been
established on photometric occasions.  (Datasets from nights when
science targets {\it only\/} were observed, with no fundamental
photometric standard-field observations, were also treated as
non-photometric.)  Color-transformation coefficients can be
derived for non-photometric datasets just as for photometric ones,
provided that stars of a range of color are available in at least
some individual images; this was always the case here.  Of the 84
datasets containing our observations of M4, 37 were considered to
be of photometric quality, and the remaining 47 were treated as
non-photometric.  

All the CCD images were reduced by PBS using the
DAOPHOT/ALLSTAR/ALLFRAME suite of programs.  These data were then
calibrated to the Johnson {\it UBV\/}, Kron/Cousins {\it RI\/}
photometric system of \citet{stetson2000,stetson2005}\footnote{
http://www1.cadc-ccda.hia-iha.nrc-cnrc.gc.ca/community/STETSON/standards/,\\
\hbox{see also}
http://www.cadc.hia-iha.nrc.gc.ca/community/STETSON/homogeneous/archive/}, 
which is designed to be as close as possible to that of 
\citet{landolt1992}; see also \citet{landolt1973,landolt1983}.  
The photometric standard values that were employed
here were those current as of 2013 August 13.  By comparison with
Landolt's published values for stars in common, we determine that
this photometric system agrees with Landolt's with an
rms dispersion {\it per star\/} of 0.028~mag in $U$,
0.018~mag in $B$, 0.013~mag in $V$, 0.011~mag in $R$, and
0.015~mag in $I$, based upon 203, 325, 339, 216, and 230
individual stars common to the two samples, respectively.  We
believe that these numbers represent a fundamental limit to the
accuracy with which these broad-band photometric indices for any given
star can be transformed from one filter set to another, given the
variety present among stellar spectral-energy distributions.  {\it
In the mean\/}, our system appears to agree with Landolt's to
$\ltsim 0.001$~mag (68\% confidence interval) in {\it BVRI\/} and
to $\ltsim 0.002$~mag in $U$ for stars in the magnitude range of
overlap ($7.6 \leq V \leq 16.1$). The equivalence of the two photometric 
systems at fainter magnitudes relies primarily on the reliability of 
the shutter timings in the various cameras used for our observations; 
we fully expect these to be accurate to well under 1\%.

For service as a local standard in the M4 field, suitable for use
in the final transformation of all instrumental magnitudes to the
adopted standard photometric system, we identified by eye stars
that appeared to be relatively free from blending with neighboring
stars.  We further required that each star have been observed in a
given filter on at least three photometric occasions, have a
derived standard error of the mean magnitude in that filter $\leq
0.04$~mag, and have no evidence of intrinsic variability in excess
of 0.05~mag, root-mean-square, based upon the repeatability of the
available observations in all filters.  These conditions were
satisfied by 634, 875, 878, 308, and 877 stars in $U$, $B$, $V$,
$R$, and $I$, respectively.  (For comparison, our usual stricter
selection criteria---specifically, at least {\it five\/}
independent observations on photometric occasions and standard
errors $< 0.02$~mag in at least {\it two\/} filters, as well as
repeatability better than 0.05~mag, r.m.s.---were satisfied by
251, 856, 860, 298, and 857 of these stars, respectively.) These
stars were then used as a local reference for the photometric
calibration of all optical measurements of stars in M4 from the final
ALLFRAME reduction.  This analysis resulted in 64,626 stars in the
M4 field with calibrated photometry in $V$ {\it and\/} at least
one of $B$ or $I$; 62,943 stars had calibrated photometry in all
three of $B$, $V$, and $I$; and 18,218 of them had calibrated
photometry in all five of $U$, $B$, $V$, $R$, and $I$.  

Since any given CCD image spans only a fraction of the total field
covered by all our observations, no star appears in every image. 
In fact, the {\it maximum\/} number of calibrated magnitude
measurements for any given star was 12 in $U$; 1,110 in $B$; 1,510
in $V$; 1,817 in $R$; and 42 in $I$.  Considering only stars with at
least one observation in a given filter, the median number of
observations was 4 in $U$, 21 in $B$, 31 in $V$, 1770 in $R$, and
14 in $I$.  Since most observations were centered near the
cluster, member stars preferentially have more than the median
number of observations; stars in the outer part of the field 
where smaller numbers of observations had been made were more
likely to be field stars.  Indeed, for the stars that we considered
to be variable candidates in the study below, the median
number of observations was 9 in $U$, 1,106 in $B$, 1,495 in $V$,
1,813 in $R$, and 30 in $I$.

An additional 36,207 stars were included in the ALLFRAME
reductions, but they fell outside the area where adequate
observations exist to define local photometric standards; for
these stars, astrometric information is available, but no
calibrated photometry is possible at present.

\section{Near infrared photometry}\label{chapt_obs_nir}

The infrared photometry for M4 consisted of 55 datasets from 18 observing
runs, as outlined in Table~\ref{tab:table_obs_ir}.  As with the optical data, all
photometric measurements were carried out by PBS using the
DAOPHOT/ALLSTAR/ALLFRAME suite of programs.  Calibration of the
data was slightly different from the optical case, in that we are
able to use the point-source catalog of the Two Micron All Sky
Survey (2MASS)\footnote{The 2MASS Point Source Catalog was
produced by a joint project of the University of Massachusetts and
the Infrared Processing and Analysis Center, funded by the
National Aeronautics and Space Administration and the National
Science Foundation.}, which provides calibrated {\it JHK\/}
photometry for stars in virtually all our images.  This gives
us the option of applying the ``non-photometric'' calibration
approach described in the previous section to obtain fundamentally
calibrated magnitudes on a well-defined system, without the need
to consider extinction corrections among observations obtained at
various airmasses.  

For our purposes we selected only those published 2MASS magnitudes
that were flagged as of photometric quality ``A,'' implying an estimated
standard error of the adopted magnitude $\sigma \leq 0.1086$.  A total of
11,919 2MASS stars having a quality ``A'' magnitude in at least one
of the three filters could be identified with entries in our own
catalog of stars in the target field within a one-arsecond match-up
tolerance, after modest astrometric transformations had reduced
the r.m.s.\ astrometric differences from 0\Sec28 in the RA
direction and 0\Sec21 in the Dec direction to 0\Sec19 in
each coordinate.  

With these stars initially adopted as photometric standards,
robust fitting techniques \citep[see][]{stetson89} were used to estimate
photometric transformations from the instrumental system of each
of our infrared datasets to the photometric system of the 2MASS
catalog.  The transformations involved linear color
transformations, photometric zero points, and linear zero-point
gradients in $x$ and $y$ across the face of each detector: 

$$j({\rm observed}) = J(\hbox{\rm standard}) + z + ax + by + c(\jmk) $$

\noindent etc.  Average values of $a,b,c$ were computed for each
detector chip and each observing run, and a separate value of $z$
was obtained for each digital image.  

There were two exceptions to this approach.  First, run~1
(designated ``m4ir'' in Table~\ref{tab:table_obs_ir}) 
seemed to be of such good quality that we adopted the
full photometric-night reduction scheme, including explicit
extinction corrections and field-to-field photometric tie-in. 
This helped to ensure a uniform photometric calibration among the
different images, which were individually rather small and
contained relatively few 2MASS stars apiece.  Second, the images
obtained with MAD---the multi-conjugate adaptive-optics 
demonstrator on the VLT \citep{march2006}---were so deep, so
small, and so close to the cluster center that they contained no
unsaturated images of high-quality 2MASS stars.  In order to
calibrate these data, we used our measurements of fainter stars
contained within the other datasets, calibrated to the 2MASS
system, to serve as local secondary standards for the MAD data.  

After an initial calibration was found for all of the datasets
excepting ``mad3,'' the transformations were applied to our
instrumental magnitudes for the 2MASS standards, weighted averages
were taken, and the results compared to the 2MASS magnitudes for
the same stars.  Those showing the largest residuals were rejected
from consideration as local standards, and new photometric
transformations were derived from all the datasets employing the
more restricted set of local 2MASS standards.  This was repeated
until all stars showing residuals larger than 0.30\,mag in any of
the three filters had been eliminated.  At the end, our
photometric calibration relies on a total of 7,901 2MASS stars
with valid measurements in $J$; 5,847 in $H$; and 3,389 in $K$. 
Computed values of the zero-point gradients ($a,b$) ranged between
extreme values of --0.151 and +0.048$\,$mag per 1000 pixels, and
the standard deviation among all the chips and runs was about
0.015$\,$mag per 1000 pixels.  The local 2MASS standards permitted
us to measure these gradients with a typical precision of
0.002$\,$mag per 1000 pixels.  The color-transformation
coefficients $c$ took on values ranging from --0.122 to +0.225,
depending upon filter and camera, with a standard deviation among
the chips/runs of about 0.16; the 2MASS stars determined these
coefficients with a typical precision of 0.008$\,$mag/mag.  The
median precision for the adopted zero-point, $z$, of an individual
exposure was 0.008$\,$mag.

The root-mean-square residual between the published 2MASS
magnitudes and our averaged calibrated photometry for the same stars is
measured to be 0.063, 0.064, and 0.069$\,$mag per star in $J$,
$H$, and $K$.  We take these as upper limits for the statistical
reliability of our average magnitude for any given star in any
given filter, since uncertainties in the 2MASS magnitudes as well
as our own contribute to the observed scatter, and the 2MASS
sample includes stars whose 2MASS standard errors are estimated to
be as large as 0.11$\,$mag.  Dividing these r.m.s.\ residuals by
the square root of $N$ suggests that the formal standard error in
transforming our instrumental magnitudes to the 2MASS system
should be $\sim 0.001\,$ in the mean, but surely this is
unrealistic:  unknown systematics could be considerably larger than
this.  For instance, we are not in a position to judge whether
there may be zone errors in the 2MASS Point Source Catalog, such
that this particular patch of sky including the globular cluster
M4 may be on a slightly different photometric system than other
parts of the sky in the same catalog.  

As mentioned above, the MAD data were calibrated using a
secondary set of local standards that we ourselves have selected
and referred to the 2MASS system.  We defined 90 of these local
standards, and there were a total of 11,844 individual magnitude
measurements for these 90 stars among the 152 MAD images.  These had
observed standard errors of 0.073$\,$mag per measurement, or
0.021$\,$mag, r.m.s., per star, between the average of the MAD
magnitudes and the average of our magnitudes for the same stars
from the other runs.  

\section{Variable Stars}

\subsection{Identification}\label{chapt_rrind}

Our optical photometry has several advantages when compared with
the various sets of time-series data available in the literature. 
(a)~In the present case, the number of consecutive observations on
any given single night could be as large as 125 in any one filter,
and up to 270 observations in all filters.  This means that we
have, for a significant fraction of objects, a number of
measurements in some filters on consecutive nights that ranges
from several tens to several hundred.  (b)~The photometric data
span a large time interval (16 years).  This allows for very
accurate period determinations and the opportunity to identify
variables with periods close to 0.5 days.   (c)~The phase coverage
is quite good in at least three photometric bands, namely $B$, $V$
and $R$. This also means that we can use three different colors
and brightness amplitudes to constrain the stars' pulsation properties
(see \S \ref{chapt_position}).  (d)~Our photometry of at least
some of the known RR~Lyrae variables can be supplemented with
photoelectric photometry avialable in the literature
\citep{sturch1977,cacc1979,liujanes1990,clementini1994}.  This
will allow us to further constrain their pulsation properties
(Blazhko effect, amplitude modulation), and gives us a chance, at
least, to look for evolutionary effects (period changes).  

We carried out independent identification of candidate variable
stars using the Welch \& Stetson index (WS,
\citet{welchstet,stetson96}) as applied to our optical data set. 
In particular, we considered all stars with a weight (defined as
the number of consecutive image pairs, plus one-half the number of
unpaired singleton images in which the star appears) greater than
$w = 180.$  We designated as candidates stars having a WS index
larger than 0.8; 242 stars met this criterion, representing 0.9\%
of all stars with $w\geq180$.  A number of the variable stars
listed in Clement's catalog failed to meet these criteria:  V43
had no data at all, falling outside the field spanned by our images;
stars V3, V29, V32, V33, V34, V42, V75, V76, and V79 had
insufficient weight, falling in the less-well-observed outer parts
of the cluster; and V17, V44, V45, V46, V47, V48, V50, V51,
V53, V54, V55, V56, V57, V58, V59, V60, V65, V71, V78 and V80 had
WS indices $< 0.8$.  We added these stars to the list of candidate
variables by hand, so that our software would extract their
light-curve data.  We will discuss these stars individually now.  

\subsection{Comparison with Clement's calatog}\label{subsec_clement}

Christine Clement's on-line catalog (``Updated 2009s'') lists 79
stars in M4 identified as variable (they are numbered V1 through
V80, but V62 is identified as equal to V55).  Of these 79 stars,
67 are listed with equatorial coordinates having a precision of
1\sec\ or better;  nine have $(x,y)$ coordinates with a
precision of 0\Sec1; and three have references to finding charts only,
with no positions given.  

For the 67 Clement stars with equatorial coordinates, we
transformed her positions to $(x,y)$ in our coordinate system, and
cross-matched with our photometric catalog for M4 without regard
to whether our stars appeared variable.  With simple translations
in $x$ and $y$, we were able to match 66 of the 67 stars within a
tolerance of 2\sec, with offsets of --0\Sec2 and --0\Sec5
in $x$ and $y$, and a star-to-star dispersion of
0\Sec5 in each coordinate.  Four-constant and six-constant
linear transformations did not improve the agreement.  The star that
could not be matched to any star in our photometry catalog was
Clement's V43; her published position for this star falls outside
the area covered by our images.  Of the 66
matching stars, 58 of the matches were unique: no other star in
our catalog was present within 2\sec\ of Clement's positions (as
transformed to our reference frame).  

In each of the other eight cases, there were two stars in our
catalog within 2\sec\ of the Clement position: V21, V40, V65,
V69, V70, V72, V73, and V78.  Among these, V21 was not ambiguous:
the star closer to the predicted position had the right apparent
magnitude and color to be an RR~Lyrae and had strong evidence of
variability in our data; the other star was 2.8~mag fainter and
showed no evidence of intrinsic variation.  In the case of V40,
both matches had roughly suitable magnitudes and colors, and both
showed strong evidence of variability; we retain both stars,
identifying the closer and brighter match with Clement's V40 and
designating the fainter and more distant one new variable
candidate C1.  Among the other six cases, there were only two
where one star of the pair showed evidence of variability.  For
V69 (this star is also identified as K50), a $V\,$=$\,$17.3 star
1\Sec9 from the predicted position showed strong evidence of
variability while a $V\,$=$\,$16.7 star 0\Sec4 from the predicted
position did not.  In the case of V73 (this star is also identified as K54), a $V\,$=$\,$17.8 star 1\Sec8
from the predicted position showed mild evidence of variability
while a $V\,$=$\,$19.9 star
0\Sec6 from the position did not.  For these two cases, we
have provisionally identified Clement's star with the one in our
data that showed evidence of variation despite the fact that it
was farther from the predicted position than the alternative.  

In the final four cases (V65, V70, V72, and V78), neither star
appeared to be variable in our data.  However, for V65, V70, and
V78 we were able to make an unambiguous match by comparing our
measured mean magnitudes for these stars to those given in
Clement's table.  V72 is the lone remaining star.  Clement gives
$B_{\hbox{\footnotesize max}} = 16.34$ for this star, with an amplitude in
$B$ of 0.3 mag.  Our two stars within 2\sec\ of the predicted
position have $\left< B \right> = 19.17$ and 18.42.  But
2\Sec01 from the predicted position there does lie a star with
$\left< B \right> = 16.46$ and strong evidence of variability. 
Accordingly we adopt this star as the match.  Alerted by this, we
checked the apparent magnitudes of the other, seemingly unambigous
matches, and found that V67 (= K48), V68 (= K49), V71 (= K52),
and V73 (= K54) had probably been misidentified; in each case another
star more than 2\Sec0 but less than 2\Sec5 from the predicted
position appeared to be a better photometric match, and
accordingly these changes were made.  

The star V17 turned out to be a special case.  Clement gives
equatorial coordinates for this star, and there was exactly one
star in our sample within 2\Sec0 of this position; in
fact, after the modest astrometric transformation that we have
applied, our star is 0\Sec31 from the position recorded by
Clement.  However, Clement gives the mean $B$-band magnitude of
V17 as 13.55, while our star has $B\,=\,14.19$ and a \bmv\ color
of 1.30, which is much too red for an RR~Lyrae.  Furthermore,
Clement says that the period of this star is 0.8555 days---which
is unusually long for an RR~Lyrae---while our star shows no
evidence of variability.  However, we note that at the reported position of
Clement's star V52 we do find an RR~Lyrae with a period of 0.8555
days (\S4.3 below). According to Clement, V17 has equatorial
coordinates 16\hr23\tm34\Ts02 --26\deg31\min07\Sec8 while
she gives the position of V52 as 16\hr23\tm24\Ts08 --26\deg30\min27\Sec6,
suggesting that at some point a transcription error changed the
right ascension of V52 by 10 seconds of time, possibly creating a
fictitious V17.  We further note that according to Clement the
coordinates of variable V18 are 16\hr23\tm34\Ts69 --26\deg31\min03\Sec9,
which might possibly account for the recorded declination of V17
assuming that someone made a slip of one line in reading a table. 
This attempted reconstruction of history is purely conjectural. 
All that we can state with certainty is (a)~at Clement's recorded
position for V17---which she attributes to a 2009 private
communication from Samus---we find no RR~Lyrae at all, let alone
an RR~Lyrae with the unusual period of 0.8555$\,$d; (b)~at
Clement's recorded position for V52 there is an RR~Lyrae with the
period of 0.8555$\,$d (she lists 0.4605$\,$d as the period of V52,
while we find {\it no\/} RR~Lyrae in M4 with a period between
0.455 and 0.463$\,$d).  For our purposes here, we drop Clement's
V17 from the list of candidate variable stars in M4 and retain
the designation V52, but we revise the star's period from 0.4605 to
0.8555$\,$d.  

The nine stars V53 through V61 have only $(x,y)$ positions
(referred to the nominal cluster center) in the Clement catalog. 
These are all supposed to be fairly luminous stars, so it was easy
to plot their predicted positions on our stacked image of the
cluster field and identify the closest bright star to each
position.  Quantitatively, the match required position offsets of
1\Sec7 in each coordinate; the nine stars then matched nine
bright stars in our catalog with a root-mean-square dispersion $<\,$0\Sec1.  

There are three variable candidates in Clement's list with no
positions given.  

According to Clement, V75 is identified as star
G343 in a finding chart published in (according to Clement)
Greenstein (1939 ApJ 90, 401) but the correct citation appears to
be Greenstein (1939 ApJ 90, 387: \cite{green39}); the relevant chart itself
appears as Plate~V in that paper.  We have identified this star in
our image of the field.  The star does not show strong evidence of
variation in our data, but the number of our observations is
comparatively small.  

V79 is also supposed to appear in
Greenstein's chart as star G302.  A dot representing that star is
not actually visible either in the on-line reproductions of
Greenstein's chart, or in the photographic reproduction of his
plate in the paper copy of the journal.  However, in our
digital image of the field, there is a blended pair of stars
between stars G299 and G303, at a position where it would lie just
to the left of the label ``302'' in Greenstein's plate.  Of these
two, the southeastern star has strong evidence of intrinsic
variability in our data; we identify this as V79.

According to Clement, V76 is equal to star
ZB14, identified by $(x,y)$ position and finding chart in
\citep{yao77}.  We have examined this paper to the extent
possible, given that none of us can read Chinese.  We find
equatorial coordinates and finding charts for two stars, ZB12 and
ZB13, but nothing that appears to be a position for ZB14.  There
is a plate in the paper, Plate 2, whose caption does not mention
ZB14 in characters that we are able to read, but there is only one
star marked in this plate.  We have indentified this same star in
our data, and find that it has quite strong evidence of intrinsic
variability on a timescale appropriate for an RR Lyrae variable,
although the number of our measurements of this star is
comparatively small.  Furthermore, the measurements
that we do have for this star imply a magnitude and colors
appropriate for an RR Lyrae at the cluster distance.  On the
working hypothesis that this is Yao's star ZB14, we provisionally
identify it as V76.  We remind the reader that we have no direct
evidence that Plate 2 in the Yao paper is intended to be a finding
chart for ZB14; this is purely conjecture on our part.  

\begin{figure*}[t]
\centering
\includegraphics[width=16cm]{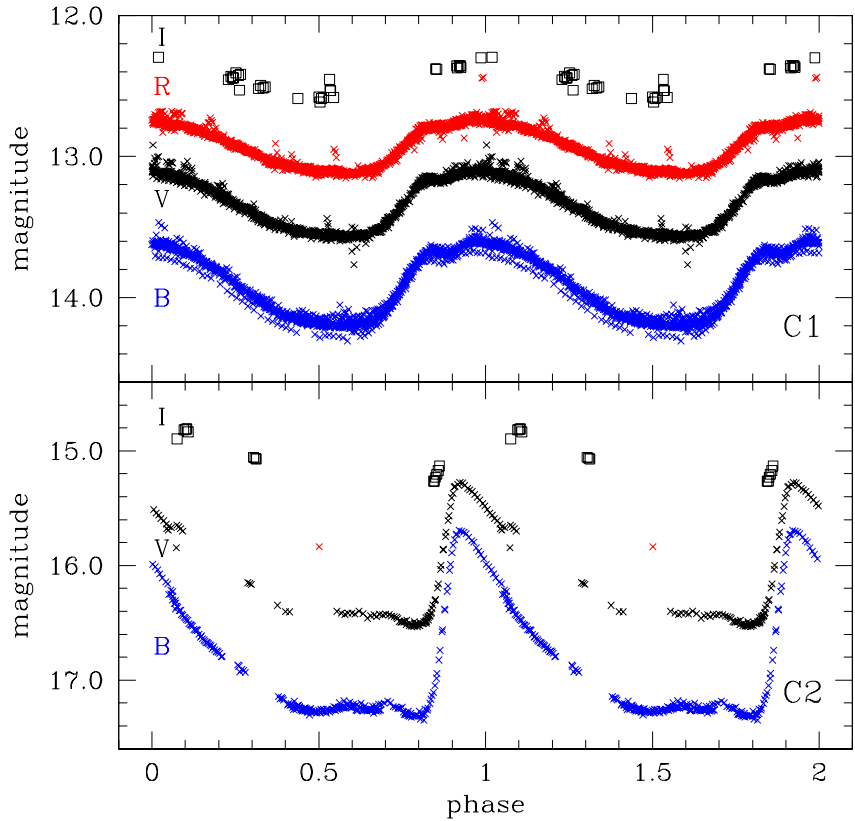}
\vspace*{1.0truecm}
\caption{Optical light curves for newly identified RR Lyrae
candidates C1 (upper) and C2 (lower).  The blue, black, and
red points represent observed magnitudes in the $B$, $V$, and $R$
photometric bands, respectively.  Empty black squares are measurements
in the $I$ band.  The stars' estimated periods are:  (C1) 0.2863$\,$d and
(C2) 0.4548$\,$d.
}
\label{fig:c1}
\end{figure*}

Among the other stars that we identified in our catalog with WS$\,>\,$0.8
and $wt \geq 180$ most could be attributed to bad data of one sort or another.
There was one other star, however, whose magnitudes could be phased into a
good RR~Lyrae light curve.  We designate this star C2.  Fig.~\ref{fig:c1}
shows the optical-band light curves for the newly identified RR~Lyrae
candidates C1 and C2.

We list our equatorial positions for the stars we have identified
as RR~Lyrae variables in the M4 field in Table~\ref{tab:pos_rr},
while  Tables~\ref{tab:phot_rr_opt} and \ref{tab:phot_rr_ir} give
photometric parameters in, respectively, the optical and
near-infrared bands for the RR~Lyrae stars.  ``V'' star names are
the identifications assigned by Clement; as already mentioned, we
have assigned ``C'' names to the new candidate variables
identified here for the first time.  For the convenience of the
reader, in Tables~\ref{tab:phot_rr_opt} and \ref{tab:phot_rr_ir}
we also provide ``K'' designations from the studies of
Kaluzny et~al.\ (1997, 2013a and 2013b).

We include V76 in these tables because in our limited data
for the star, we see short-term (i.e., over the course of an hour
or so) increases and decreases in brightness, parallel in $B$ and
$V$, and consistent with the rates expected for an RR Lyrae. 
However, our data are not extensive enough to obtain a period. 

As already mentioned, Clement's catalog also identifies a number
of candidate variable stars that are apparently not of the RR
Lyrae type.  Table~\ref{tab:table_candopt} gives our recovered
positions, average optical photometry, and a comment for each of
these, while Table~\ref{tab:table_candir} gives average NIR
photometry for the same variable candidates.  In addition, we have
looked for ten more variable candidates from \cite{kaluz2013b}
that were not included in Clement's compilation, K58--66 and K68,
and one more from \cite{kaluz2013a}, K69.  We used their tabulated
coordinates for these stars to establish their approximate
locations on our stacked image of the field, and then compared our
image to their postage-stamp finding charts to confirm
that the correct stars had been identified.  We were unable to
recover their star K58; if their tables and chart are correct,
then K58 is a $V\approx20.4$ object closely blended with a
$V=16.38$ star in our data, and we did not distinguish it as a
separate object.  Finally, we include five more new candidate
variables that we have identified for the first time here
(C3--C7).  

\begin{figure*}[t]
\centering
\includegraphics[width=14cm]{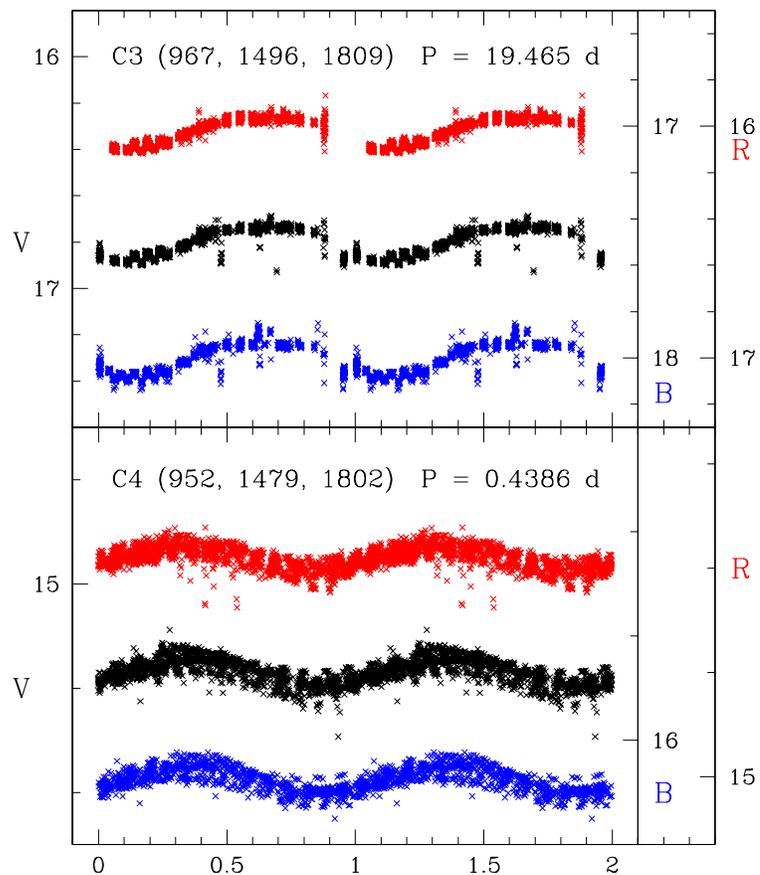}
\vspace*{1.0truecm}
\caption{Optical light curves for newly identified variable
candidates C3 (upper) and C4 (lower).  The blue, black, and
red points represent observed magnitudes in the $B$, $V$, and $R$
photometric bands, respectively.  Corresponding ordinate axes are
labeled in the same colors.  The legend gives the number of
observations in each filteri and the period for each star.  Tick
marks in the lower plot are separated by 0.20$\,$mag.
}
\label{fig:c3}
\end{figure*}

\begin{figure*}[t]
\centering
\includegraphics[width=14cm]{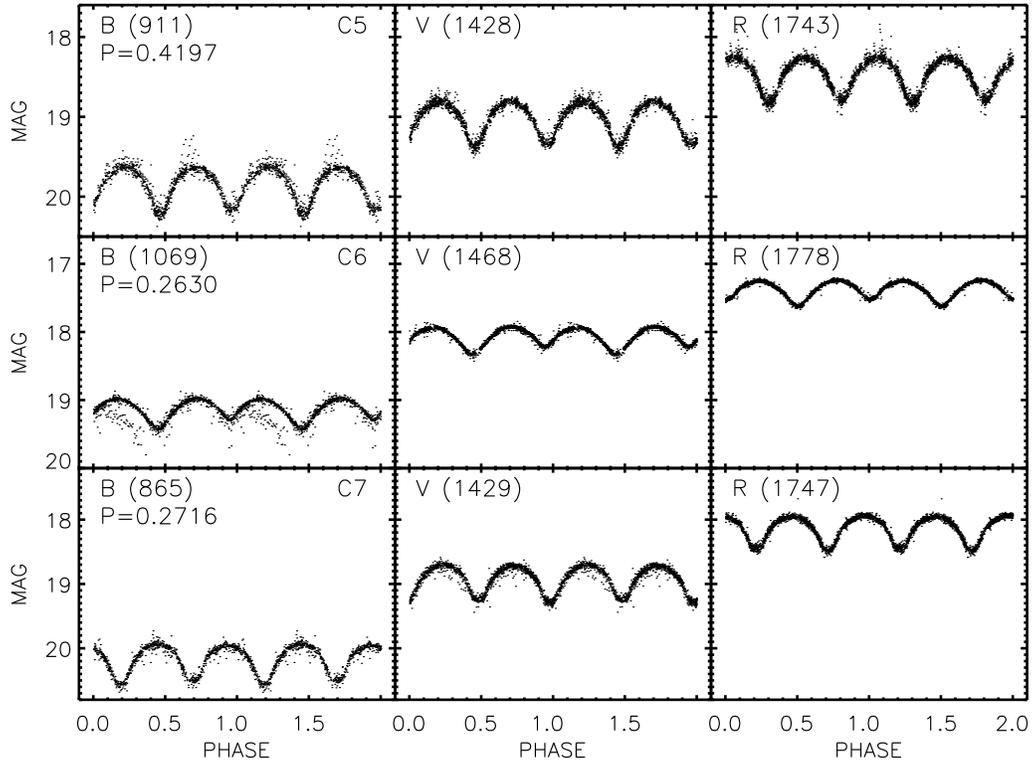}
\vspace*{1.0truecm}
\caption{Optical light curves for the newly identified
variable candidates C5 (top), C6 (middle), and C7 (bottom) in
the $B$ (left), $V$ (middle), and $R$ (right) photometric bandpasses.
}
\label{fig:large_m4_c5_7_final}
\end{figure*}

Among these stars, V53 shows some evidence for variability on
timescales of minutes to hours, but we are not able to come up
with a phased light curve or period.  Stars V54--V60 and also V65
(=$\,$K46) and K63 have been called ``LPV or EB?''
because our data show some (in some cases slight) evidence of
different mean magnitudes in at least one of $B$, $V$, or $R$ from
different observing runs, but no evidence of coherent variation
{\it within\/} any single observing run.  Conversely, of course,
it is possible that some of these stars are subject to crowding or
other photometric defects.  We would not have dared to identify
any of them as variable candidates on the strength of our own
data, had they not already been so designated in the literature. 
V77 (=$\,$K56) is mostly constant in brightness, but on Julian day
2451702 it does show one decrease and increase of about
0.06$\,$mag in the $B$ bandpass that looks very much like the
bottom of an eclipse.  For V78 (=$\,$K57) we have computed nightly
average magnitudes in $B$, $V$, and $R$; there is one night where
it is 0.04 mag fainter in $B$ than average of the 45 other nights
when the star was observed in the $B$ filter, and one night where
it is 0.08 mag fainter in $V$ than the average of the 44 other
nights when $V$ was observed, but they are not the same night, nor
are they from the same run.  There is no other evidence than this
in our data that the star we have identified as V78 is variable. 
We also have fairly extensive data for V80, and do not see
evidence for variability on time scales of either hours, days, or
years.  The other candidates, for all of which we list periods in
Table~\ref{tab:table_candopt}, are almost certainly eclipsing
binaries except perhaps C3.  We are not sure of its
classification, but with a sinusoidal light curve, small
amplitude, and a provisional period of 19 days (and it is possible
that this is an alias of the true period) it could perhaps be a
background semiregular variable.  In Figs.~\ref{fig:c3} and
\ref{fig:large_m4_c5_7_final} we illustrate the light curves of
the new variables C3--C7.  

\begin{figure*}[tbp]
\centering
\includegraphics[width=15cm]{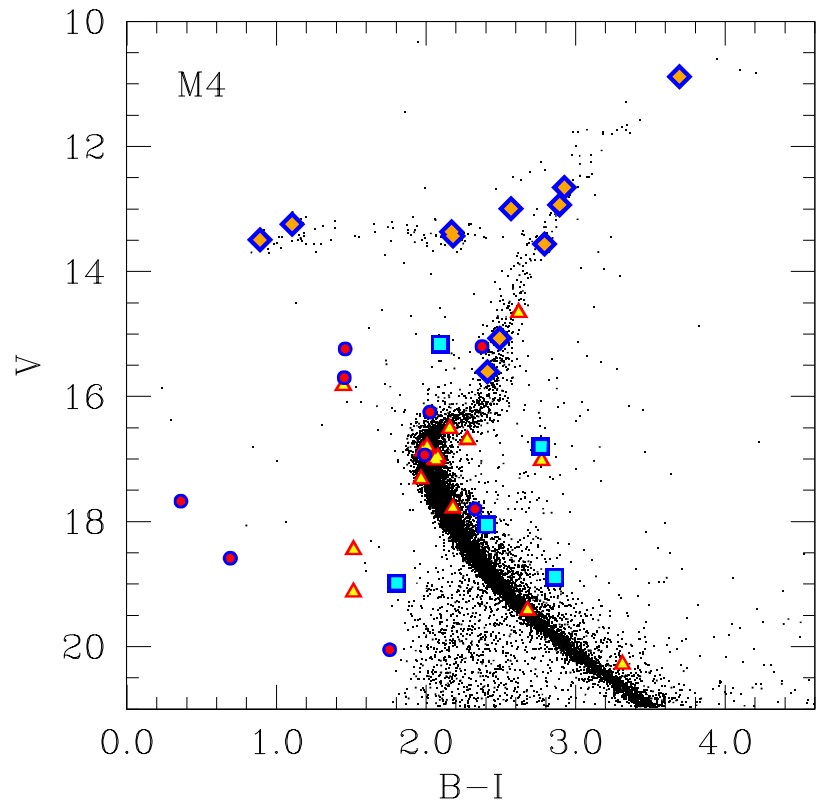}
\vspace*{1.5truecm} 
\caption{A $V$ versus \bmi\ color-magnitude diagram for M4, with
the variable candidates {\it not\/} of the RR Lyrae type
illustrated as colored symbols.  The red-and-yellow triangles represent the
candidate variables of Kaluzny et~al.\ (1997, 2013a and 2013b) for
which we are able to recover the published periods from our data.  The
blue-and-red circles are for Kaluzny stars whose variability we
are unable to confirm.  The blue-and-cyan squares
represent the five newly discovered candidates C3 through C7.  Finally, the
large blue-and-orange lozenges are for the remaining stars without
Kaluzny designations (V53--60, V75, V79 and V80) where we are
unable to confirm the variability.} 
\label{fig:nonRRcmd}
\end{figure*}

In Fig.~\ref{fig:nonRRcmd} we show the positions of all the
non-RR~Lyrae variable candidates in the color-magnitude diagram of
M4.  For this diagram, we adopt \bmi\ as the color because several
of the candidates are missing $R$-band magnitudes, but all have
$I$.  Note that two of our newly discovered candidates (C6 at
$V$, \bmi$\,=\,$18, 2.4 and C7 at 19, 2.9)
lie on M4's binary main sequence; they are thus
good candidates for cluster membership.  
These stars will not be discussed further in this paper.  

\subsection{Characterization of the variability}

A period search was performed for all candidates using a robust
string-length algorithm that we have developed 
\citep{stetson96,stetson98}.  This identifies the best candidate
period from simultaneous consideration of the calibrated measurements and
their associated standard errors in all available 
photometric bands.  (In our case, we used {\it BVRI\/}; $U$ was omitted because of
the comparatively small number and relatively poor quality of measurements.  It must
be pointed out that the $I$-band data contributed relatively little weight to this
procedure.)  The graphical
display incorporated in our software makes it easy to distinguish
real periodic variability from corrupt data, and the stars
displaying the latter rather than the former were manually triaged
out of our sample at this point.  The type of variability (e.g.,
fundamental-mode RR~Lyrae, first overtone RR~Lyrae, Blazhko effect; eclipsing
binary; long-period variable) is also fairly evident at this point
and we were able to enter this information in our notes.  Once the
best candidate period had been identified, Fourier series were
then fitted by nonlinear least-squares to refine the candidate
periods and estimate the mean magnitudes and amplitudes in the
different photometric bands.  

These initial estimates of the variables' properties were then
independently confirmed or challenged by different co-authors
employing two other algorithms: 

(1)~In our variant of the Lomb-Scargle method \citep{scargle82}
algorithm, the candidate period is identified employing
simultaneously all the measurements available in the different
photometric bands ({\it UBVR\/}). A graphical interface allows fine
tuning of the candidate period.    

(2)~Our new version of the PDM method \citep{stellingw} identifies
the period using only the measurements in the photometric band
with the largest number of measurements and/or the largest time interval 
covered.  

\noindent For this stage of the analysis, our own measurements of the M4
variable candidates were supplemented by whatever time-stamped
observations were to be found in literature.  All our initial
classifications of the reality and type of variability were
unchanged by this confirmation/challenge step.  The difference in
the periods measured by the above methods are typically smaller
than one part per one hundred thousand. Once we had agreed upon
the period, the individual measurements were phased and we
performed a fit of the {\it UBVR\/} light curves using splines
under tension. Some researchers feel that this method may be more
robust for objects that show either noisy data or amplitude
modulation. These analytical fits allowed us to reliably estimate
the mean magnitudes, the brightness amplitudes, and the epochs of
maximum light.  

The numbers of measurements available in the $I$ band are too
limited to perform analytical fits with the above methods.
Therefore, having adopted the $I$-band template light curves for
RR~Lyrae stars that have been provided by \citet{lay98} together
with our new estimates of period, $V$-band amplitude and epoch of
maximum, we performed a least-squares fit to estimate the mean
$I$-band magnitude and amplitude for each star.  Template light
curves for the RR~Lyrae stars in the $U$ band are not yet
available.  For this reason, we have provisionally estimated the
mean $U$-band magnitudes of the RR~Lyraes either from the unweighted
mean of individual measurements (29 stars) or with a spline fit
(18 stars); obviously, these mean $U$-band
magnitudes should be regarded as suggestive only, and scientific
use should be made of them only with extreme caution.

Table \ref{tab:phot_rr_opt} gives for each star, from left to
right, the identification, the period based on the LS method, the
mean {\it UBVR\/} band magnitudes based on spline fits, the
mean $I$-band magnitude based on the template fit, and the epoch of maximum light.  We also give
the $U$-, $B$-, $V$-, $R$- (spline fit) and $I$-band (template fit)
amplitudes.  For completeness, we have included V34, V43 and V76 in these tables
despite the fact that we do not have independent photometric data for
V43, and we are unable to perform a proper independent light-curve
analysis from our skimpy data for V34 and V76.  

As discussed above, we have cross-correlated our list of candidate
variable stars with the list of variable and candidate variable
stars provided by \citet{clement2001}.  We confirm variability for
44 of her 45 RR Lyrae candidates (excluding V43, for which we have
no photometric data of our own, but including V34 and V76, for
which we have enough data to see significant brightness changes,
but not enough to fit a light curve) and provide updated estimates
of their periods for 43 (excluding V34 and V76, but including V43,
for which we use published data).  Our sample also includes the
two newly discovered highly probable RR~Lyraes (C1 and C2).  In
total, if we include the published photometry for V43 and the
published periods for V34 and V76, there are 45 cluster
RR~Lyraes and two field RR~Lyraes (C2 and V64) with
periods that we believe are reliable, and photometry that we
believe is useful.  

\begin{figure*}[tbp]
\centering
\includegraphics[width=14cm,height=14cm]{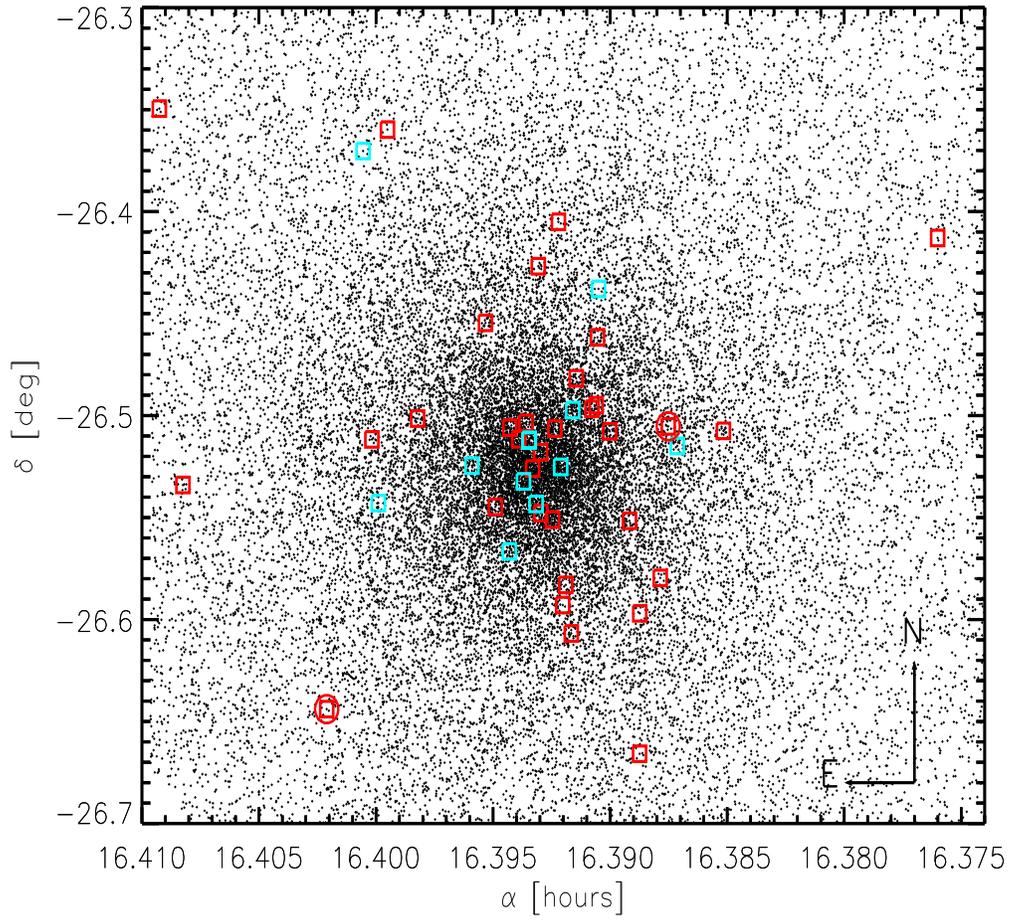}
\caption{Distribution on the sky of RR Lyrae stars in the M4 field
for which we have accurate optical and astrometric measurements.
The sky area covered by our photometry is 35\ \farcm$\times$35\
\farcm. The red and cyan squares show the positions of
fundamental and first overtone variables.  Two probable field
stars are circled.  Note that eclipsing binaries, candidate
long-period variables, and other variable types are not marked
here.} 
\label{fig:fig_radec}
\end{figure*}

Fig.~\ref{fig:fig_radec} shows the positions on the sky of both
the cluster and field RR~Lyraes.  The red squares show the positions
of fundamental-mode pulsators, while cyan squares are for
first-overtone pulsators.  The positions of the two candidate
field RR~Lyrae stars are additionally marked by enclosing circles.  
A stacked digital image of our field in M4 and a list of the pixel
coordinates of all the variable candidates can be found at our web 
site\footnote{\bf http://www.cadc-ccda.hia-iha.nrc-cnrc.gc.ca/community/STETSON/homogeneous/NGC6121}.

\section{The RR~Lyraes in the Color-Magnitude Diagram}\label{chapt_position}

\begin{figure*}[tbp]
\includegraphics[width=14cm,height=14cm]{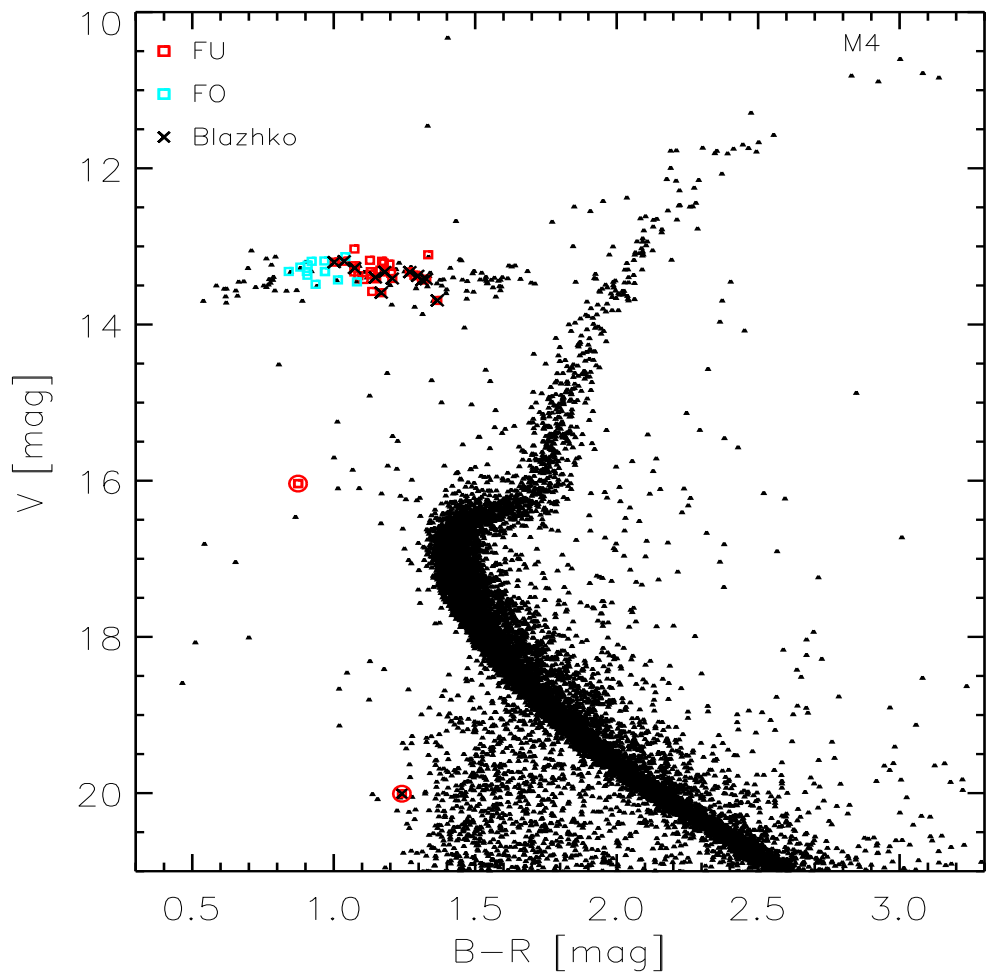}
\centering
\caption{
Optical $(V,\bmr)$ color-magnitude diagram for the M4 field. RR~Lyrae variables 
pulsating in the fundamental (FU) or in the first overtone (FO) are represented 
by red and cyan squares, respectively.  Black crosses identify
candidate Blazhko stars.  Two probable field stars are circled.
Numerous field stars with $\bmr \gtsim =1.3$ are seen below the
cluster main sequence.
}
\label{fig:fig_cmd_vbr}
\end{figure*}

\begin{figure*}[tbp]
\centering
\includegraphics[width=12cm]{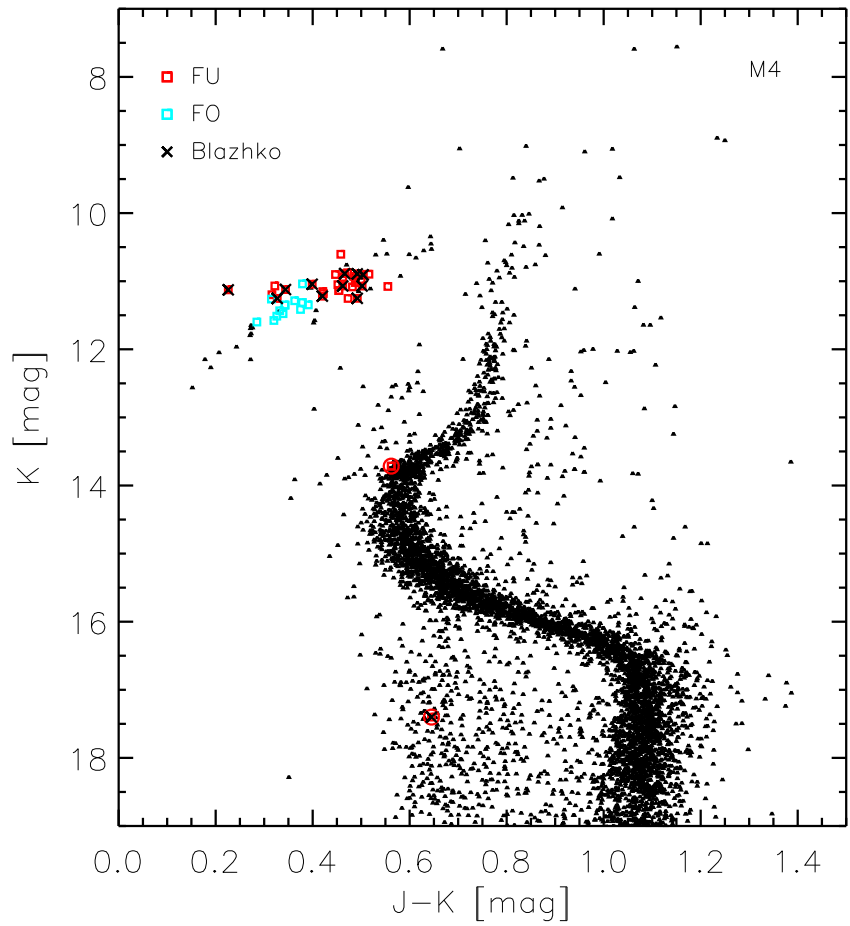}
\caption{Near-infrared $(K,\jmk)$ color-magnitude diagram for the M4 field.  RR~Lyrae variables
pulsating in the fundamental mode (FU) or in the first overtone (FO) are marked 
with red and cyan squares, respectively. The black crosses designate candidate Blazhko 
RR~Lyrae stars. The two candidate field RR~Lyrae stars are located at $K\sim13.8$ and 
$K\sim17\,$mag. The field stars show a well defined blue edge with colors ranging 
from \jmk=0.6 to \jmk=1.2.}\label{fig:fig_cmd_nir}
\end{figure*}

Fig.~\ref{fig:fig_cmd_vbr} shows the $V$, \bmr\ color-magnitude
diagram (CMD) of M4.  (We use \bmr\ here because most stars have much better
measurements in $R$ than in $I$.)  Stars were selected according to the number
of measurements ($\ge$ 10), the photometric standard error
($\sigma_{BVR}\le$ 0.02 mag), and the radial distance ($1^\prime \le r
\le 7^\prime$) from the center of the cluster (which we have
estimated to be 16\hr23\tm35\Ts21, --26\deg31\min34\Sec7, with
an estimated uncertainty $\sim\,$1\sec\ in each coordinate). 
Contamination by field stars is quite evident below the cluster
main sequence for \bmr\ colors ranging from 1.3 to 2.5. The high
number of individual measurements in {\it BVR\/} allows us to
have good photometric precision over the entire magnitude range. 
Fig.~\ref{fig:fig_cmd_nir} shows the $K$ versus \jmk\ color-magnitude diagram for
M4.

In both Fig.~\ref{fig:fig_cmd_vbr} and Fig.~\ref{fig:fig_cmd_nir} the red and cyan
symbols represent the positions of fundamental-mode (FU) and
first-overtone (FO) RR~Lyraes, respectively (see \S6 below for a more
detailed discussion concerning the identification of the pulsation
mode); the two probable field stars have circles around them and lie far from
M4's horizontal branch.  The candidate Blazhko stars are marked
with black crosses.  As expected, FOs in M4 have mean \bmr\ colors
that are systematically bluer than FUs \citep{bono97b}.  

The two candidate field RR~Lyrae stars (C2, V64) are found at
$V\sim16$ and $V\sim20\,$ mag ($K \sim 13.7$ and 17.4).   The
latter, V64, is quite interesting since its apparent magnitude places it
in the outskirts of the Galactic halo. Recent photometric
catalogs---the ASAS Survey \citep{pojm2002}, the QUEST Survey
\citep{vivas04}, the NSVS survey \citep{kine06}, the LONEOS Survey
\citep{mice08}, the Catalina Real-time Transient survey
\citep{drake09,djorg11}, the SEKBO survey \citep{akht12}, the LINEAR
Survey \citep{pala13}, the GALEX Time Domain Survey \citep{geza13} in
the ultraviolet, and the Carnegie RRL Program \citep{freed12} in
the mid-infrared---have included only a very few RR~Lyraes in these halo regions.  
C2 lies at Galactic coordinates $(l,b) = (351, +16)$, and it appears
2.7$\,$mag fainter in $V$ than the RR~Lyraes in M4 ($<V>\,=\,16.04$
versus 13.32---the median of the mean $V$ magnitudes for 45 RR~Lyraes in M4).
Thus, all other things being equal, C2 is 3.5 times farther away than M4, or
7.7$\,$kpc if we take 2.2$\,$kpc as the distance to M4.  This implies a
$Z$-distance of 2.1 kpc above the plane and a radial distance of roughly
2.5$\,$kpc from the center of the Galaxy.  This places it within the Galactic
bulge.

\section{Mode classification and pulsation properties}\label{chapt_rrlyrsample}

\begin{figure*}[tbp]
\centering
\includegraphics[width=12cm]{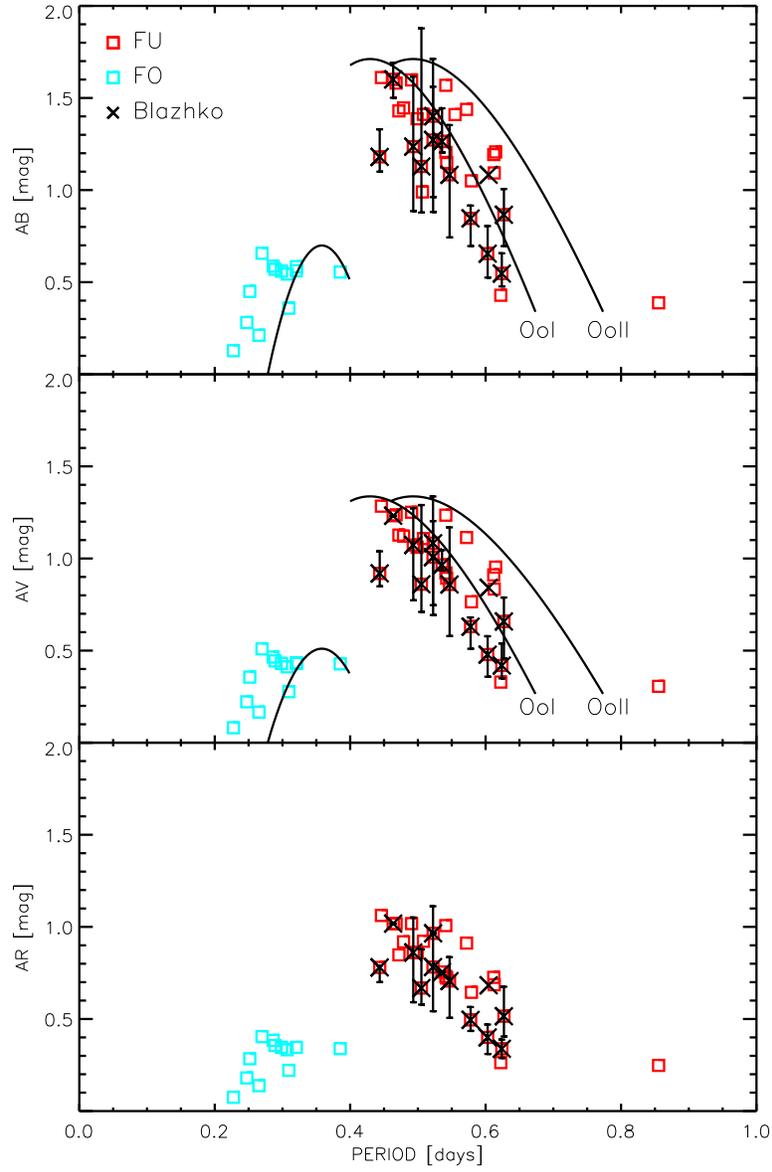}
\vspace*{1.5truecm} 
\caption{Top: A plot of the $B$-band amplitude versus period for RR~Lyrae stars in M4 (Bailey diagram). 
Fundamental and first overtone RR~Lyraes are plotted as red and cyan squares.
Candidate Blazhko stars are marked with a black cross and the vertical bars display 
their range in brightness amplitude.
Middle: Same as the top, but for the $V$-band amplitude.
Bottom: Same as the top, but for the $R$-band amplitude.}
\label{fig:bailey}
\end{figure*}

The mode identification of the RR~Lyrae stars in the field of M4
was based on the so-called Bailey diagram, i.e., the plot of brightness
amplitude versus period (Fig.~\ref{fig:bailey}). In this plane, the RR~Lyrae stars
show a clear separation between FU and FO pulsators, and are
also easily distinguished from other radial variables (Classical
Cepheids, Type II Cepheids, Anomalous Cepheids, high amplitude
delta Scuti stars).  Fig.~\ref{fig:bailey} shows that FO
pulsators (cyan squares) have---as expected---shorter periods
and smaller amplitudes than FU pulsators (red squares). The
brightness amplitudes---again as expected---decrease monotonically
when moving from the $B$ (top panel) to the $R$ (bottom panel)
photometric band. Moreover, the FU amplitudes decrease steadily
with increasing period. This trend has been attributed to the
increasing efficiency of convective energy transport when moving
from the hot to the cool edge of the FU-mode instability strip
\citep{bonostell1994,bono97b}.  The two labeled solid curves in the right side of the diagram
represent the mean loci of FU variables of Oosterhoff classes I and II,
as defined by \cite{cacc2005}.

The candidate Blazkho RR~Lyraes (black crosses) are all pulsating
in the FU mode and they cover the entire period range of FU
pulsators.  The fraction of Blazhko stars is roughly 40\% (13 out of
31), but recent space and ground measurements indicate that the
fraction of RR~Lyraes showing the Blazhko phenomenon is more
typically of the order of 50\%
\citep{benko2010,jurc2009,sodor2012,kunder2013b}. The vertical bars
attached to the Blazhko crosses in Fig.~\ref{fig:bailey}
represent the modulation of the amplitude according to our
photometry. They are preliminary estimates, since we have not
performed a detailed analysis of the secondary modulation; better
time series data than we presently have would be more suitable for
that purpose.  However, our data do indicate that the Blazhko
amplitude generally decreases when moving from the hot to the cool edge of
the fundamental-mode instability strip.  In addition, we see a half dozen FU
variables that, at fixed period, have larger brightness
amplitudes ($A_B$, $A_V$) than the other RR~Lyraes (i.e., placing them above/to the right
of the OoI locus in the figure), including two of the
Blazkho stars.  This suggests a possible systematic difference
between these stars and the bulk of the  variables.  We will find
in the next section that a group of variables showing similar
properties is also present in M5.

Variable V52 is characterized by very low amplitudes and a period
(P=0.8555$\,$d) significantly longer than the other low-amplitude
FU pulsators. It is located close to the red edge of the
instability strip (\bmr=1.33 mag), and appears to be 0.20 mag
brighter in $V$ than the other RR~Lyraes with similar \bmr\
colors. This all suggests that V52 is well evolved from the zero-age
horizontal branch.       

In general, the FO pulsators have periods shorter than 0.4 days
and amplitudes showing the typical bell \citep{bono97b}---also
sometimes called hairpin \citep{kunder2013c}---distribution
represented by the solid curve in the left side of the Bailey
diagram. The complete separation between FO and FU pulsators in
this diagram is what allows us to classify them unambiguously. 
The new candidate member variable has a position in this diagram
that is consistent with the previously known candidates.  Three of
the FO pulsators, including the new one, have very short periods
and small amplitudes, suggesting that they are located close to
the blue edge of the first overtone instability strip.    

The Bailey diagram indicates that the RR~Lyraes in M4 cover the
entire width of the instability strip, but so far we lack any
evidence of the presence of mixed-mode pulsators \citep{kunder2011}.

\begin{figure*}[t]
\centering
\includegraphics[width=14cm]{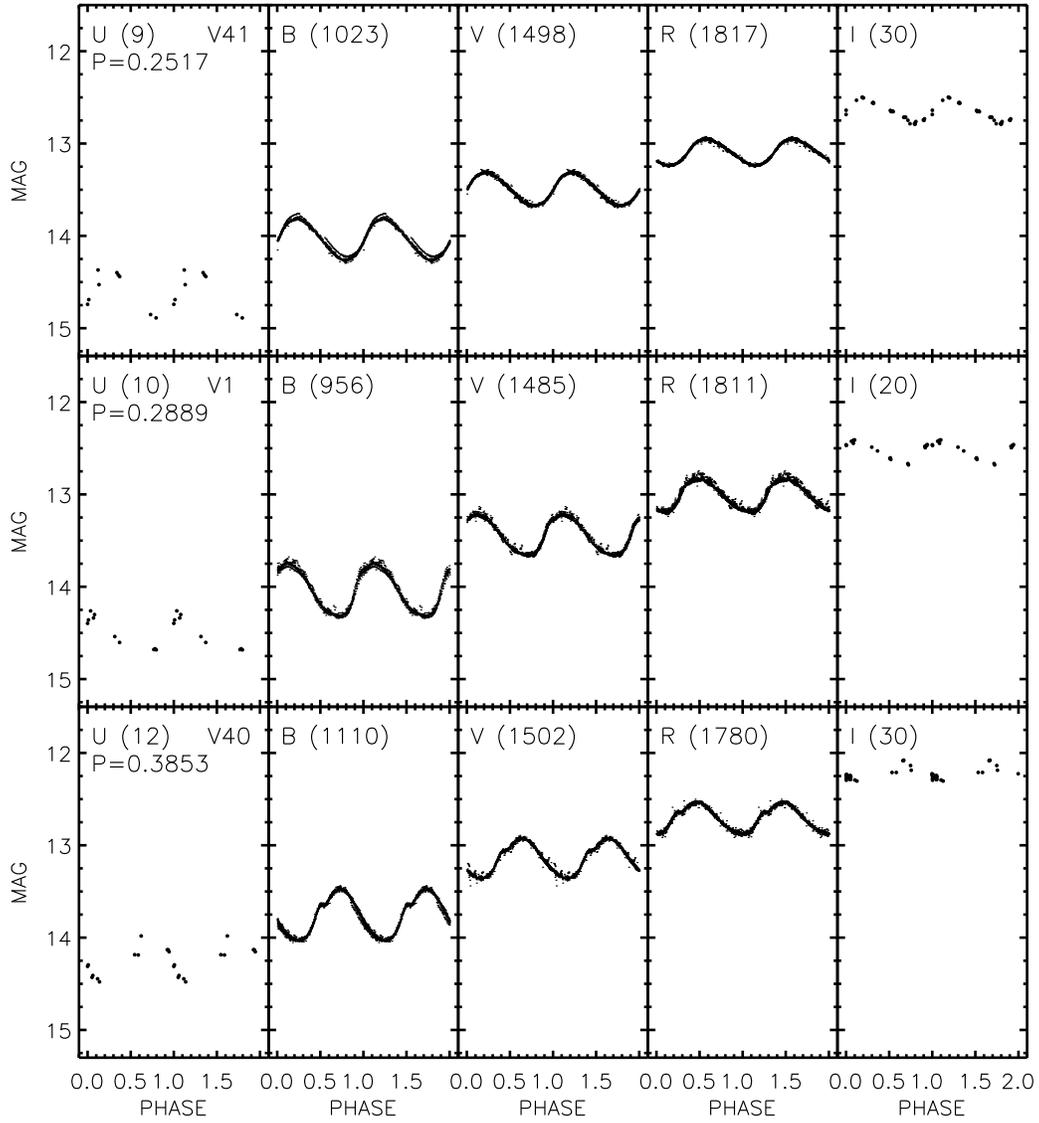}
\vspace*{1.0truecm}
\caption{Optical light curves for some of the first overtone RR~Lyrae stars in M4.
From top to bottom the panels display the light curves of
different variables ordered by increasing period.  From left to
right the different panels display the light curves in the
$U,\,B,\,V,\,R$ and $I$ bands.  The numbers in parentheses give
the number of individual measurements per band. 
The identification numbers of the variables are given in the top
right corner of the leftmost panel.  The complete atlas of FO
light curves is available in the electronic edition.}
\label{fig:large_fo_1_paper}
\end{figure*}

\begin{figure*}[t]
\centering
\includegraphics[width=14cm]{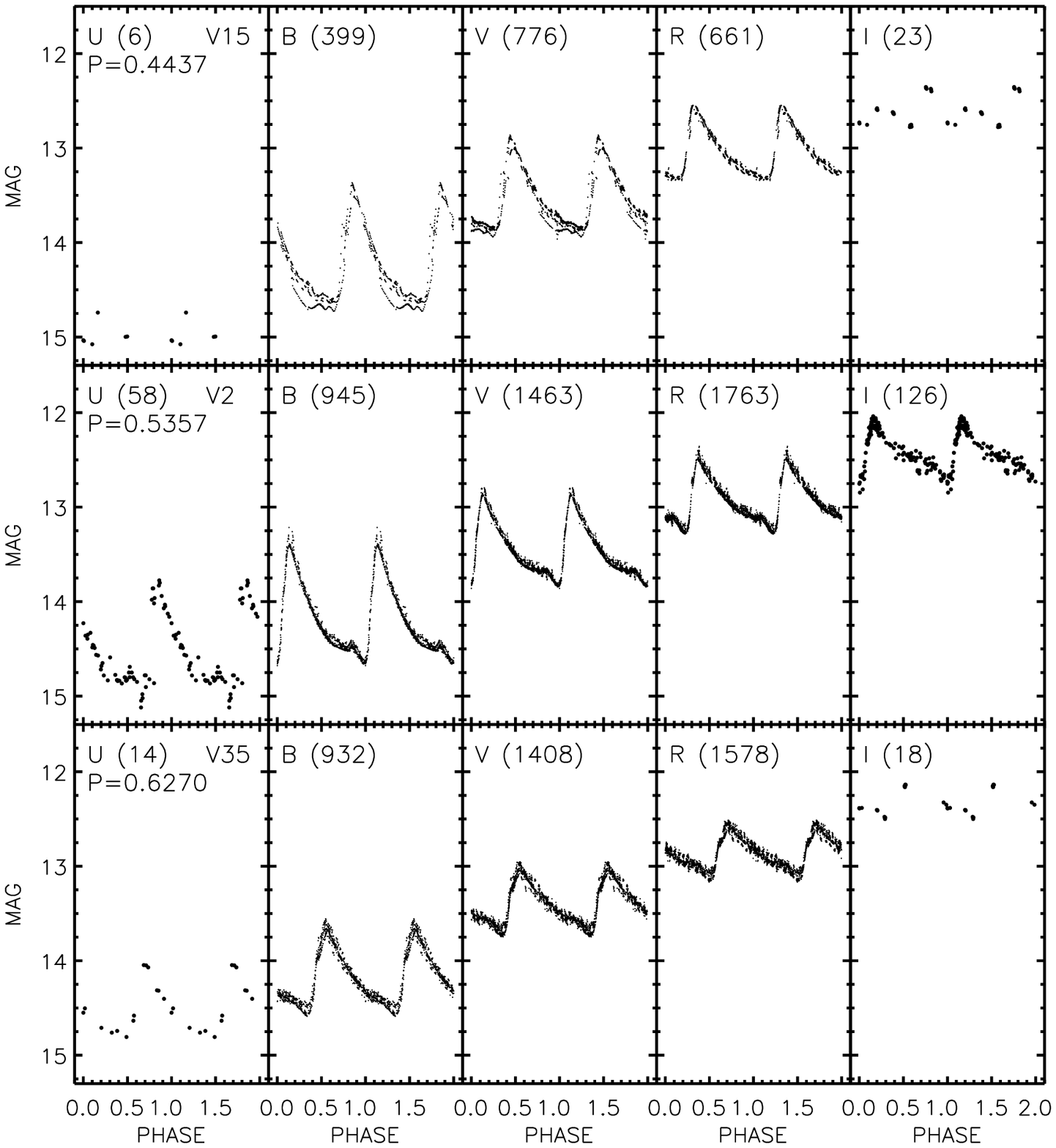}
\vspace*{1.0truecm}
\caption{Same as Fig.~\ref{fig:large_fo_1_paper}, but for fundamental pulsators.
The complete atlas of the FU light curves is available in the 
electronic edition.}
\label{fig:large_fu_1_paper}
\end{figure*}

\begin{figure*}[t]
\centering
\includegraphics[width=12cm,height=12cm]{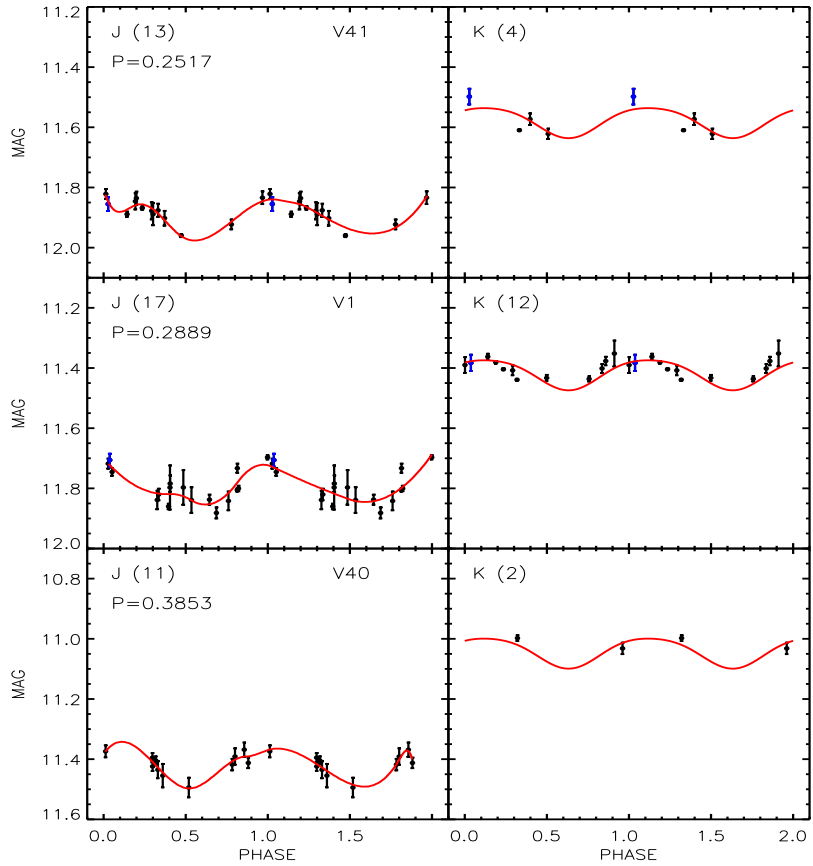}
\vspace*{1.0truecm}
\caption{Near-infrared light curves for the same first overtone (FO) RR~Lyrae stars as in Fig.~\ref{fig:large_fo_1_paper}.
From top to bottom, the panels display the
light curves of different variables ordered by increasing period.
The left and the right panels display the light curves in the $J$ 
and $K$ bands. The numbers in parentheses report the number of individual
measurements per band. The blue dots show the phase points based 
on 2MASS photometry. The vertical error bars display the standard 
deviation of the averaged phase points. The solid red curves plotted 
in the left panels show the spline fit, while the red curves in the 
right panels show the fitted template light curves (see text for
more details).  
The complete atlas of FO NIR light curves is available in the electronic 
edition.}
\label{fig:large_fo_1_ir_paper}
\end{figure*}

\begin{figure*}[t]
\centering
\includegraphics[width=12cm]{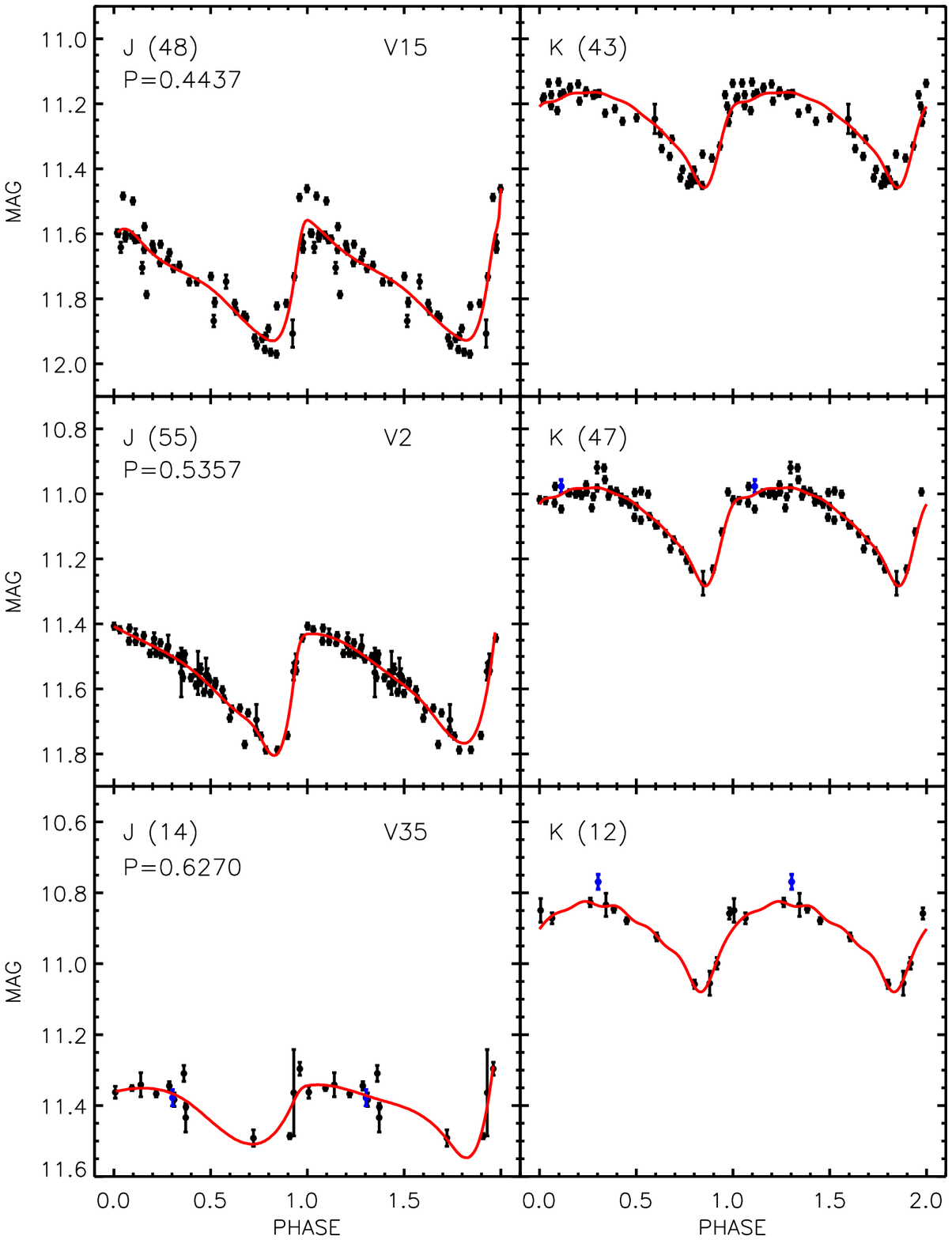}
\vspace*{1.0truecm}
\caption{Same as Fig.~\ref{fig:large_fo_1_ir_paper}, but for fundamental pulsators.
The complete atlas of the FU NIR light curves is available in the 
electronic edition.
}
\label{fig:large_fu_1_ir_paper}
\end{figure*}

Fig.~\ref{fig:large_fo_1_paper} displays the light curves of a representative
sample of first-overtone 
pulsators, while Fig.~\ref{fig:large_fu_1_paper} shows the light curves of  
FU pulsators. From top to bottom, the RR~Lyraes are ranked 
by increasing pulsation period; from left to right the different panels 
display the light curves in the $U$, $B$, $V$, $R$ and $I$ bands. The
numbers in parentheses specify the number of points plotted, 
while IDs of the variables and the pulsation periods are given in
the leftmost panel.     Figs.~\ref{fig:large_fo_1_ir_paper} and
\ref{fig:large_fu_1_ir_paper} illustrate our near-infrared {\it JHK\/}
data for the same stars.

\begin{figure*}[t]
\centering
\includegraphics[width=12cm]{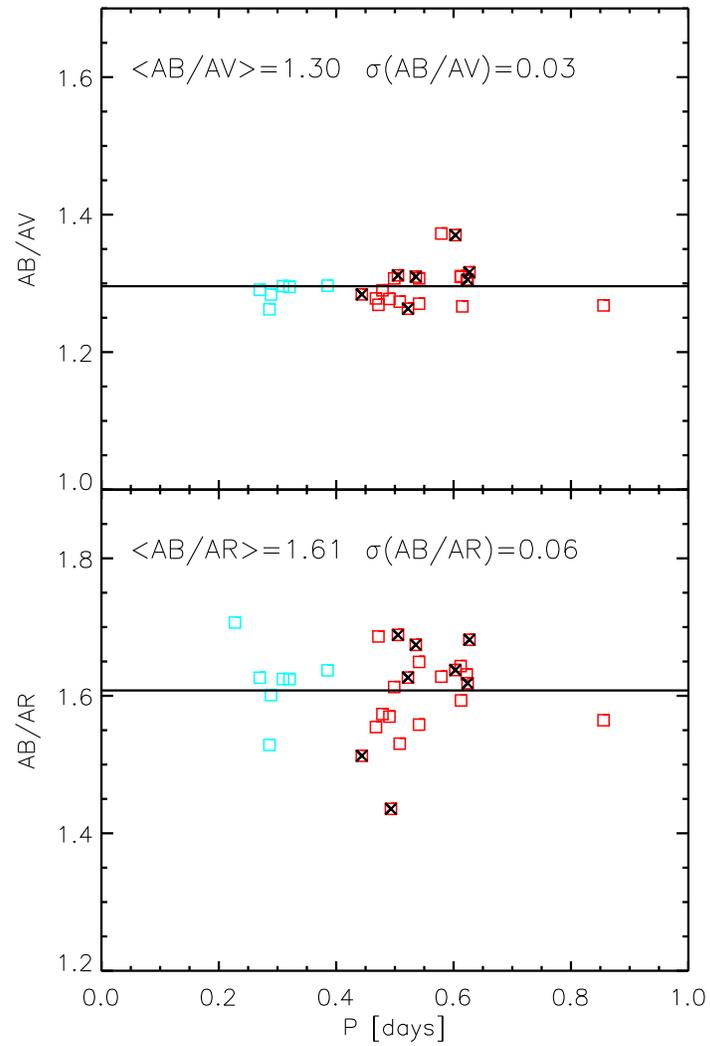}
\vspace*{1.0truecm}
\\
\caption{Top: Ratio between $B$- and $V$-band amplitudes for FU (red squares) 
and FO (cyan squares) variables as a function of period. The black
crosses mark candidate Blazhko RR~Lyraes.  The horizontal black line shows
the mean value of the ratio.  
Bottom: Same as the top, but for the $B$- versus $R$-band amplitude ratio. 
}
\label{fig:Amplitude_ratio_veryg_opt_final}
\end{figure*}

We have also investigated the ratios between the brightness
amplitudes as measured in the different optical bands. This
diagnostic serves a double purpose:  objects that show clear
discrepancies in these ratios are likely to be either (a)~other
types of variable or (b)~blended with other stars.  For instance, theoretical
pulsation models and empirical data indicate that the optical
amplitude ratios of FU and FO Cepheids should be similar
\citep{szab2012}, but the same is not true for RR~Lyraes
\citep{kunder2013c}. The data plotted in the top panel of
Fig.~\ref{fig:Amplitude_ratio_veryg_opt_final} show that the ratio
between the $B$- and $V$-band amplitudes in M4 maintains a quite
stable value over the entire period range of the FU and FO
variables. A similar result is found for the $B$- to $R$-band
amplitude ratios (bottom panel), but here the spread is slightly
larger (0.06 or 3.8\% versus 0.03 or 2.3\%). The lack of clear outliers is additional
evidence of the ``clean'' nature of the RR~Lyrae sample in M4.  

\begin{figure*}[t]
\centering
\includegraphics[width=14cm]{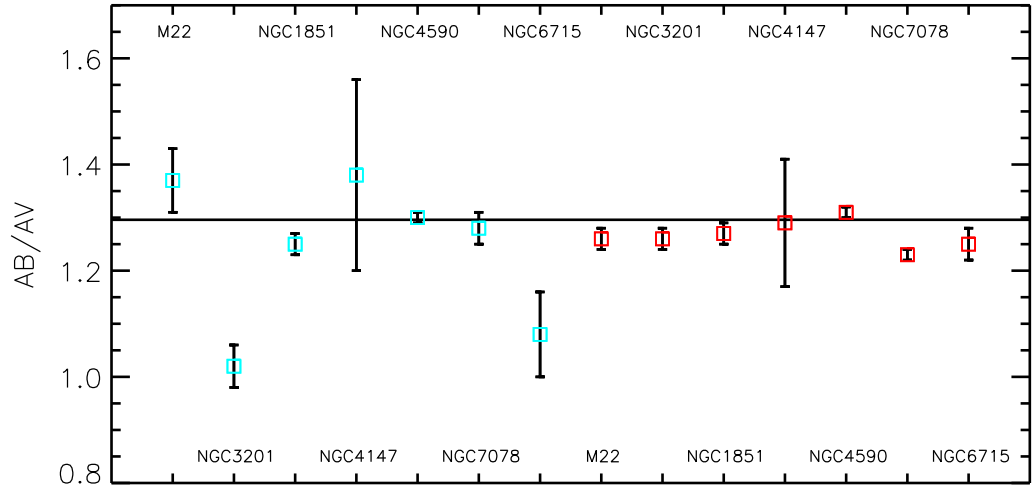}
\vspace*{1.0truecm}
\\
\caption{Mean values of the $B$- and $V$-band amplitude ratio for RR~Lyrae stars
in the sample of globulars collected by \citet{kunder2013c} (see their 
Table~3 and 4). The mean amplitude ratios for FO (cyan squares) and FU (red squares)
and their error bars as given by Kunder et al.\ are plotted at arbitrary periods.   The
horizontal solid line indicates the value of this ratio that we have measured in M4.
}
\label{fig:fig_amplratio_cluster}
\end{figure*}

Moreover, a comparison of the mean amplitude ratios in M4 with
similar ratios for seven other Galactic globulars collected by
\cite{kunder2013c} shows very good agreement (Fig.~\ref{fig:fig_amplratio_cluster}), within the errors,
for both FO (blue squares) and FU (red squares) pulsators. Three
out of the seven globulars are OoII clusters (\ngc{7078},
\ngc{4590}, M22) and the remaining four are OoI clusters
(\ngc{3201}, \ngc{1851}, \ngc{4147}, \ngc{6715}).  Note that the
two outliers, \ngc{3201} and \ngc{6715}, are both OoI clusters, 
but each has only a limited sample of FO variables: four and eight respectively. 
However, these Oosterhoff classifications should be treated with
caution because both M4 and \ngc{6715} could be considered
Oosterhooff-intermediate, depending upon the classification diagnostic
(either the mean period, or the (RRc):(RRab+RRc) population ratio) one chooses to
adopt.  In particular, the mean period of the entire sample of
RR~Lyraes in \ngc{6715} is 0.60$\,$d,  i.e., intermediate between OoI
and OoII, while for M4 it is 0.55$\,$d, i.e., typical for OoI
clusters. On the other hand, the population ratio for \ngc{6715}
is 0.14 \citep{sollima2010}, i.e., typical for OoI
clusters, while for M4 it is 0.30, i.e., intermediate between OoI
and OoII clusters.    

\section{The RR~Lyraes in M4 and M5}\label{chapt_m4m5}

To better interpret the pulsation properties of RR~Lyraes in M4 we
have made a detailed comparison with the RR~Lyraes in
M5$\,=\,$\ngc{5904}. Accurate iron abundances based on
high-resolution spectra give similar results for the two clusters:
[Fe/H]=--1.17$\pm$0.05 (estimated systematic uncertainty) for M4 and --1.34$\pm$0.06 for M5
\citep{carr09}.  Note that the mean iron abundance
of M4 has been solidly confirmed by \cite{malavolta2014} on
the basis of medium-resolution spectra for $\sim$2,800 stars collected with
FLAMES/GIRAFFE at VLT. Specifically, they found
$<$[Fe/H]$>$=--1.07, $\sigma$=0.025 (standard deviation, one star) for RGB stars and
\hbox{$<$[Fe/H]$>$=--1.16}, $\sigma$=0.09 for subgiant and main-sequence
stars.  The available accurate spectroscopic measurements  also
indicate similar $\alpha$-element abundances in M4 and M5
\citep{pritzl2004}.  Moreover, M5 is also affected by very low
reddening (E(\bmv)=0.03, Harris 1996) and essentially no
differential reddening, increasing its value as a reference
standard. 

   It should also be noted that precise spectroscopic analysis has
   suggested that there are some differences in detail between the
   heavy-element mixes of the two clusters.  For instance,
   \cite{yong2008} state, among other conclusions, ``\ldots, (2)~the
   elements from Ca to Ni have indistinguishable compositions in M4
   and M5, (3)~Si, Cu, Zn, and all $s$-process elements are
   approximately 0.3$\,$dex overabundant in M4 relative to M5, and
   (4)~the $r$-process elements Sm, Eu, Gd, and Th are slightly
   overabundant in M5 relative to M4.''  However, the astrophysical
   significance of these minor differences---what they mean for the
   physical state of the environment in which the clusters formed,
   and how they have affected the evolutionary processes that occurred in
   the individual stars of the two clusters---are not fully
   understood.  For our immediate purposes, it is sufficient 
   that all the available evidence indicates that the overall
   abundances of elements heavier than helium, the theoreticians'
   $Z$, are the same for the two clusters within well under a factor
   of 2 (0.3$\,$dex).

\begin{figure*}[tbp]
\centering
\includegraphics[width=14cm,height=18cm]{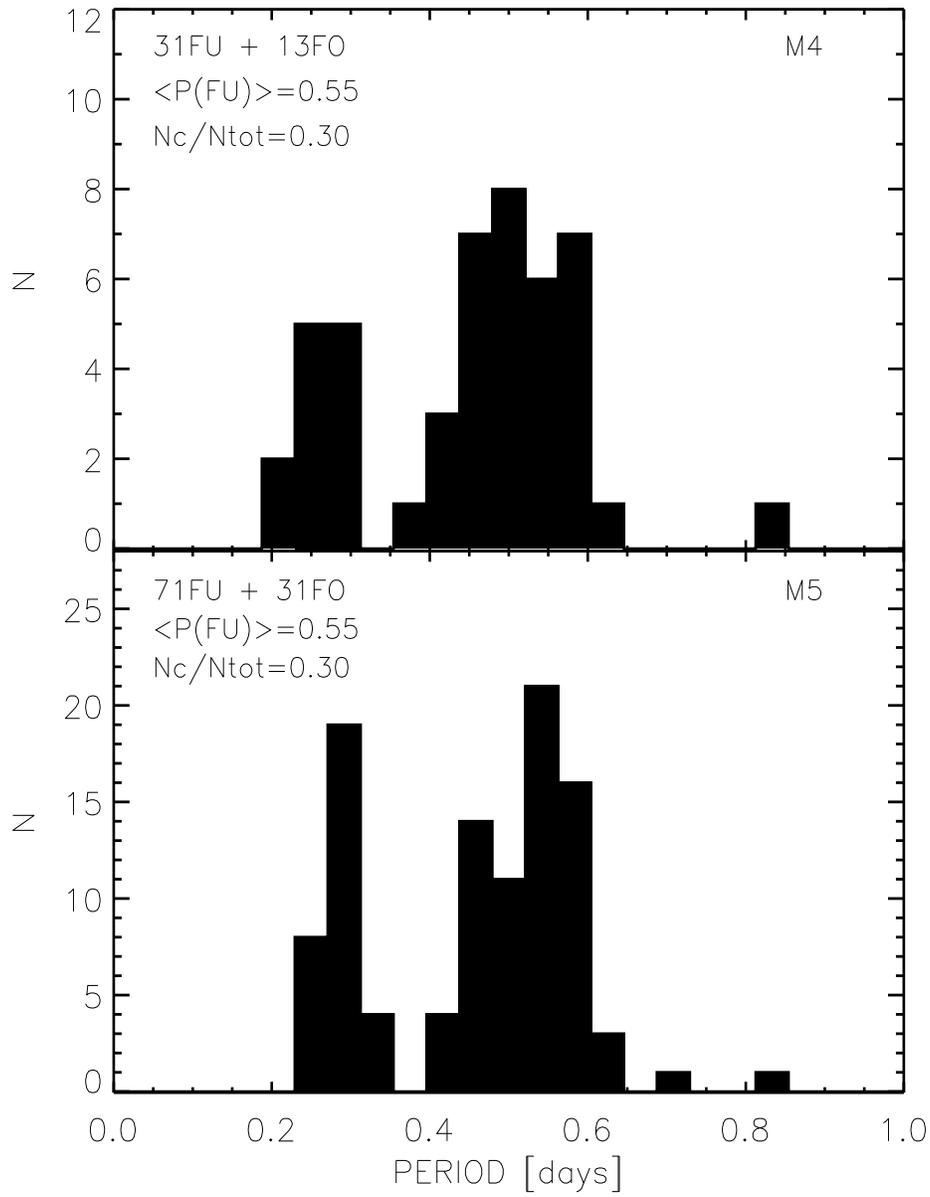}
\vspace*{1.5truecm} 
\caption{Top: Period distribution for RR~Lyrae stars in M4. The
legend gives the actual numbers of fundamental and first overtone pulsators, the mean period of the FU pulsators,
and the ratio of the number of RRc-type (FO) pulsators to the total number of RR~Lyraes (Nc/Ntot).
See text for more details.
Bottom: Same as the top, but for RR~Lyraes in M5 \citep{coppola2011}.}
\label{fig:period_final}
\end{figure*}

To quantify the comparison of the pulsation properties of the
RR~Lyraes in M4 and in M5, we considered two observables that are
by their nature unaffected by reddening and extinction: the
periods and brightness amplitudes of the variables. 
Fig.~\ref{fig:period_final} shows that the period-frequency
distributions in the two clusters are quite similar. The two
samples differ in size by slightly more than a factor of two (45
versus 102), but within the counting statistics they show very similar
period distributions. In particular, each displays two well
separated peaks: one for FO pulsators, at P$\sim$0.3$\,$d, and one
for FU, at P$\sim$0.5$\,$d.   More quantitatively, 
we find that the FU pulsators in the two clusters have identical mean periods
(0.55$\,$d).  The same conclusion comes from the RR~Lyrae population
ratios, i.e., the ratio of the number of FOs to the total number of
RR~Lyraes: we find that (RRc):(RRab+RRc) is 0.30 for both M4 and M5.  

\begin{figure*}[tbp]
\centering
\includegraphics[width=13cm]{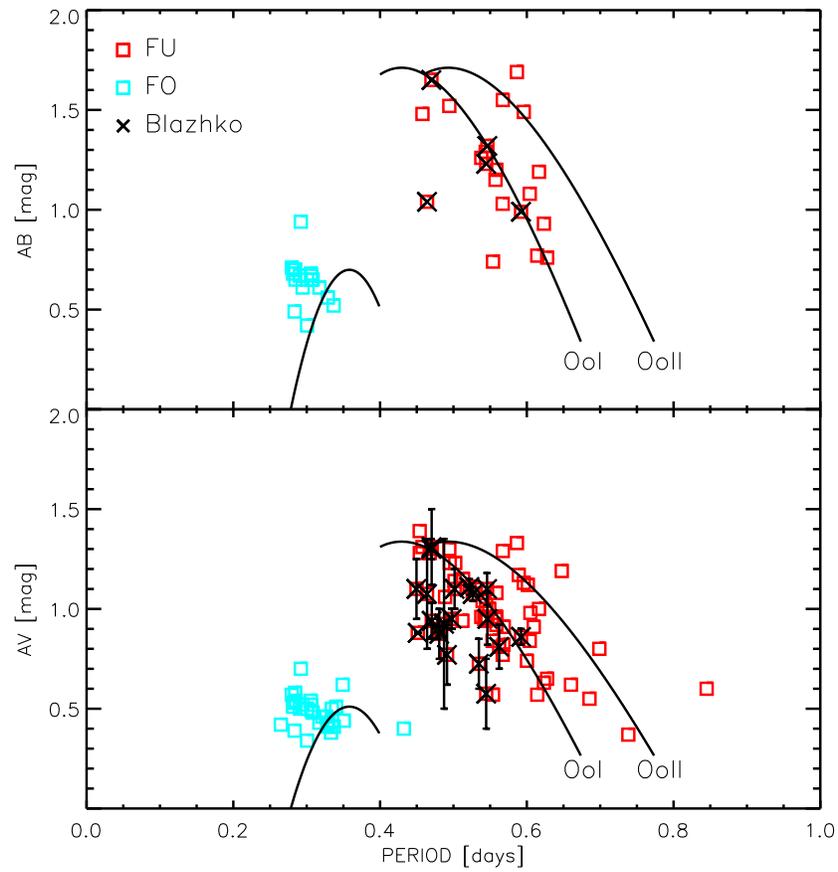}
\caption{Same as Fig.~\ref{fig:bailey}, but for RR~Lyrae stars in M5 \citep{szeidl2011}. 
Top: $B$-band amplitude versus period. RR~Lyraes pulsating in the fundamental 
and in the first overtone are plotted as red and cyan squares, respectively. 
Candidate Blazhko stars are marked with a black cross and the vertical 
bars display their range of brightness amplitude in the $V$ band; the published data
do not allow us to determine the range of brightness amplitude in $B$.
Bottom: Same as the top, but for the $V$-band amplitude.}
\label{fig:bailey_m5_Szeidl2011_final}
\end{figure*}

The Bailey diagram for M5
(Fig.~\ref{fig:bailey_m5_Szeidl2011_final}), which is inherently
reddening- and extinction-independent on both its axes, is also qualitatively
very similar to that of M4 (Fig.~\ref{fig:bailey} above).

To further quantify the comparison between the helium-burning
phases of stars in M4 and M5, we decided to extend the comparison
to the entire horizontal branch.  We adopted the HB parameter
introduced by \cite{lee94}: $\tau_{\hbox{\footnotesize
HB}}$= (B-R)/(B+V+R), where roman B, V and R indicate the number
of HB stars bluer than the RR~Lyrae stars, the number of RR~Lyraes
and the number of HB stars redder than the RR~Lyraes (not to be
confused with italic $B$, $V$ and $R$, which represent photometric
bandpasses). Recent estimates of $\tau_{HB}$ for both M4 and M5
(Castellani et~al.\ 2013, in preparation) indicate  that the HB
morphology in the above clusters is similar (0.24 versus 0.37 on a
scale of --1.0 to +1.0) and spans the entire flat region of the
HB. This indicates that red giant branch stars in the two clusters
must have experienced very similar mass loss rates.  On the other
hand, the HB morphology of the Oosterhoff-intermediate cluster
\ngc{6715}, is significantly bluer ($\tau_{\hbox{\footnotesize
HB}}$=0.87), suggesting that the mass loss rate in this slightly
more metal-poor cluster ([Fe/H]$\,\approx\,$--1.6) was more
efficient than in M4 and M5. These findings suggest that the
simultaneous comparison of observed RR~Lyrae period distributions
and HB morphologies can provide new constraints for theories of
the late stages of stellar evolution
\citep{dotter2008,vandenberg2013}.  Particularly and importantly,
we can begin to consider whether the so-called second-parameter
problem and the Oosterhoff dichotomy might be two sides of the
same coin \citep{bono97a}.  

In this context it is worth recalling once again that Galactic
globular clusters hosting sizable samples of RR~Lyraes can be
divided into two groups according to the pulsation properties of
the variables: Oosterhoff type I (OoI) clusters are characterized
by mean FU periods <P>$\,\sim\,$0.55$\,$d, and population ratios
(RRc):(RRab+RRc)$\,\sim\,$0.17, while the Oosterhoff type II (OoII)
clusters have longer mean periods, <P>$\,\sim\,$0.65, and larger FO
populations, (RRc):(RRab+RRc)$\,\sim\,$0.44.    According to this
classification both M4 and M5 appear to be OoI clusters on the
basis of the mean period and OoI-intermediate on the basis of the
pulsation-mode population ratio. It is also worth
mentioning here that the brightness amplitudes plotted in
Figs.~\ref{fig:bailey} and \ref{fig:bailey_m5_Szeidl2011_final}
support the interpretation that both M4 and M5 are OoI clusters.  The two
black lines toward the right side of each figure display the
typical loci of OoI and OoII clusters according to
\citet{cacc2005}, while the ``hairpin'' black lines toward the left side of
the figures display the mean locus for FO pulsators in 14 OoII GCs
according to \citet{kunder2013c}.  We still lack an analytical relation for FO
pulsators in OoI GCs due to their much smaller numbers, but the systematic
offset of the RRc variables in both M4 and M5 from the OoII locus
suggests, once again, a non-OoII nature for the RR~Lyraes in both clusters.  

During the last few years, evidence has mounted that GGCs for
which the census of RR~Lyrae stars is either complete or
approaching completeness display this ``neutral Oo status''
\citep{kunder2013a,kunder2013b} if the Bailey diagram, the period
distributions, and the population ratios are all taken into account
\citep{catelan10}.  

\section{Summary and Conclusions}\label{chapt_conclusion}

We have presented optical and near-infrared {\it UBVRIJHK\/}
photometry of the Galactic globular cluster M4 (\ngc{6121}), with
particular emphasis on the RR~Lyrae variable stars in the cluster. 
We have made a particular effort to accurately reidentify the
previously discovered variables, and have been fortunate to
discover two new probable RR~Lyrae variables in the M4 field: by
their positions on the sky and their photometric properties one
is a probable member of the cluster and the other is a probable
background (bulge?) object.  We conclude that Clement's star V17
is probably in fact V52 with an incorrectly reported position, and
that the previously reported period estimate for V52 was in error. 
We have also provisionally reidentified and published accurate positions for
three RR Lyrae candidates previously identified only in finding charts.

For the convenience of future researchers we have published
accurate equatorial coordinates for all 47 candidate RR~Lyrae
stars, new mean magnitudes  for 46 of them, and new period
estimates for 45.  We have also included positions and mean
photometry for 39 candidate variable stars of other types,
including 34 previously identified in the literature and five
newly discovered here.  For 18 of these we are not able to
confidently confirm variability or estimate periods; in some
cases this is due to limitations of our time sampling, in others
it is because the stars are not really variable.  For the
remaining 22, most of which are probably eclipsing binaries, we do
provide new period estimates.  

We have illustrated optical and near infrared color-magnitude
diagrams for the cluster and have shown the locations of the
variable stars in them.  We have presented the Bailey
(period-amplitude) diagrams and the period-frequency histograms
for M4 and have compared them to the corresponding diagrams for
M5.  We conclude that the RR~Lyrae populations in the two clusters
are quite similar in all the characteristic properties that we
have considered.  The mean periods, pulsation-mode ratios, and
Bailey diagrams of these two clusters show support for the
recently proposed ``Oosterhoff-neutral'' classification.  

Our calibrated photometry for M4, a finding chart, equatorial and
pixel coordinates for all stars, and light-curve data for the
variable candidates are all available from our web site.  These
astronomical data will also be the raw material for more detailed
astrophysical analyses of the physical nature of the RR~Lyraes in
M4 and of the cluster itself, in the global context of the
formation and evolution of the Galactic halo.  These papers are
currently in preparation.  

\acknowledgments
It is a pleasure to thank John Grula, the Librarian of the
Carnegie Observatories, for providing us several papers that were
not available on ADS.  This work was partially supported by
PRIN--INAF 2011 ``Tracing the formation and evolution of the
Galactic halo with VST'' (P.I.: M. Marconi) and by PRIN--MIUR
(2010LY5N2T) ``Chemical and dynamical evolution of the Milky Way
and Local Group galaxies'' (P.I.: F. Matteucci).  PBS is pleased
to acknowledge financial support from the Erasmus
Mundus-AstroMundus Consortium of Universities.  G.B. thanks The
Carnegie Observatories visitor programme for support as a science
visitor.  
 
\appendix 
\section{Notes on individual RR Lyrae stars} 

V1 -- The $B$, $V$ and $R$ light curves show a few data points out of phase. 
The mean magnitudes and the amplitudes are minimally affected.  

V2 -- Small amplitude modulations in the $B$- and $V$-band light curves indicate that V2
might be a candidate Blazhko variable. The mean $U,I$ magnitudes and their amplitudes are based on 
\citet{sturch1977}, \citet{cacc1979}, and \citet{liujanes1990} for the $U$-band data and 
on \citet{clementini1994} for the $I$-band data.

V3 -- The current photometry covers only the phases around maximum (increasing 
and decreasing branch). The $B$-band amplitude is based on data from \citep{clement2001}.  

V6 -- The mean $U$-band magnitude and amplitude are based on data from \citet{cacc1979}.

V7 -- The current $A_V$ amplitude is smaller than than previous estimates (1.06 versus 1.24).
The difference might be due to the fact that the period of V7 is close to 0.5$\,$d. 

V11 -- Amplitude estimates in all bands are affected by very large Blazhko 
modulation. The mean $U$-band magnitude and amplitude are based on data from \citet{sturch1977}.

V12 -- The mean $U$-band magnitude and amplitude are based on data from \citet{sturch1977}.

V14 -- Very small amplitude modulations only visible in the $B$-band light curve suggest that V14 
might be a Blazhko variable. The mean $U$-band magnitude and 
amplitude ($A_U$) are based on data from \citet{sturch1977} and \citet{cacc1979}.

V15 -- Amplitude modulations in the $B$-, $V$- and $R$-band light
curves indicate that V15 is a candidate Blazhko variable.
According to \citet{clementini1994}, this is a peculiar variable
with a decreasing period and either a strong Blazhko effect or a
pulsation mode switching from FU to FO. The behavior resembles the
RR~Lyrae V79 in M3 \citet{gora80}. The $B$- and $V$-band light
curves show a larger amplitude modulation than in the $R$ band.  

V18 -- We confirm that this is an FU mode RR~Lyrae.  

V19 -- The mean $U$-band magnitude and amplitude are based on data from \citet{sturch1977}.

V21 -- The current $B$-, $V$- and $R$-band amplitudes are smaller
than previous estimates, probably due to the fact that the pulsation period is close to 0.5$\,$d. 

V22 -- Amplitude modulations in the $B$- and $V$-band light curves indicate that V22 is a
candidate Blazhko variable.

V24 -- Amplitude modulations in the $B$-, $V$-- and $R$-band light curves indicate that V24 
is a candidate Blazhko variable. 

V27 -- The mean $U$-band magnitude and amplitude are based on data from \citet{sturch1977} and
\citet{cacc1979}.

V28 -- Amplitude modulations in the $B$-, $V$- and $R$-band light curves indicate that V28 
is a candidate Blazhko variable. The mean $U$-band magnitude and amplitude are based 
on data from \citet{sturch1977}.

V29 -- Small amplitude modulations in the $B$- and $V$-band light curves suggest that V29 
may be a candidate Blazhko variable. The period, the mean
magnitudes, the amplitudes and the epoch of maximum include data from
\citet{sturch1977}, \citet{cacc1979} and \citet{clementini1994}.  

V30 -- The mean $U$-band magnitude and amplitude are based on data from \citet{cacc1979}.

V31 -- Amplitude modulations in the $B$-, $V$- and $R$-band light curves indicate that V31 
is a candidate Blazhko variable. The mean $U$-band magnitude and amplitude are based 
on data from \citet{sturch1977}.

V32 -- The period, the mean magnitudes, the amplitudes and the epoch of maximum include 
data from \citet{sturch1977}, \citet{cacc1979} and \citet{liujanes1990}.

V33 -- The period, the mean magnitudes, the amplitudes and the epoch of maximum include 
data from \citet{sturch1977}, \citet{cacc1979} and \citet{liujanes1990} for the 
$U$, $B$, and $V$ bands and from \citet{liujanes1990} for the $R$ and $I$ bands.

V34 -- Limited number of data points in all bands. 

V35 -- Moderate amplitude modulations in the $B$-, $V$- and $R$-band light curves indicate that V35 
is a candidate Blazhko variable. The mean $U$-band magnitude and  
amplitude are based on data from \citet{sturch1977} and \citet{cacc1979}.

V36 -- Somewhat noisy light curves. 

V38 -- The $B$- and the $V$-band light curves show small amplitude modulations. It might 
be a plausible candidate to have been or to be becoming a Blazhko variable. 

V39 -- Amplitude modulations in the $B$-, $V$- and $R$-band light curves indicate that V39
is a candidate Blazhko variable. 

V42 -- The period, the mean magnitudes, the amplitudes and the epoch of maximum include 
data from \citet{cacc1979} and \citet{clementini1994}.

V43 -- The astrometry is from \cite{clement2001}.  The photometric parameters are based on data from 
\citet{cacc1979}.

V49 -- This is a low-amplitude FO variable with a very short
period. The light curves are noisy but its color is among the bluest of the entire sample.  

V64 -- Amplitude modulations in the $B$-, $V$- and $R$-band light curves indicate that V64 
is a candidate Blazhko variable. This is V45 in \citet{kaluz1997}.  This is clearly a distant
background object unrelated to M4.

C1 -- New cluster FO variable. 

C2 -- New candidate field FU variable, a background object unrelated to M4.  

According to our current photometric results these candidate
RR~Lyrae variables listed in the \citet{clement2001} catalog, do
not seem to be variables: V17, V44, V45, V46, V47, V48, V50, and V51.
For the candidate variable V34 we have strong evidence of RR Lyrae-like
variability, but do not have enough phase points to determine a
period or define a light curve.  For V43 we have only astrometric information and
no new calibrated photometry of our own.
\bibliographystyle{apj}

\begin{thebibliography}{100}

\bibitem[Akhter et~al.\ (2012)]{akht12} Akhter,~S., Da~Costa,~G.~S., Keller,~S.~C., \& Schmidt,~B.~P. 2012, ApJ , 756, 23


\bibitem[Bassa et~al.\ (2004)]{bassa2004} Bassa,~C. et~al.\ 2004 ApJ, 609, 755



\bibitem[Benkho et~al.\ (2010)]{benko2010} Benkho,~J.~M., et~al.\ 2010, MNRAS, 409, 1585

\bibitem[Bono et~al.\ (1997a)]{bono97a} Bono,~G., Caputo,~F., Cassisi,~S., Incerpi,~R., \& Marconi,~M. 1997, ApJ, 483, 811

\bibitem[Bono et~al.\ (1997b)]{bono97b} Bono, G., Caputo, F., Castellani, V., \& Marconi, M. 1997, A\&AS, 121, 327




\bibitem[Bono \& Stellingwerf (1994)]{bonostell1994} Bono,~G., \& Stellingwerf,~R.~F. 1994, ApJS, 93, 233


\bibitem[Cacciari (1979)]{cacc1979} Cacciari,~C. 1979, AJ, 84, 1542

\bibitem[Cacciari et~al.\ (2005)]{cacc2005} Cacciari,~C., Corwin,~T.~M., \& Carney,~B.~W. 2005, AJ, 129, 267

\bibitem[Cardelli, Clayton \& Mathis (1989)]{cardelli}Cardelli,~J.~A., Clayton,~G.~C., \& Mathis,~J.~S. 1989, IAUS, 135, 5

\bibitem[Carretta et~al.\ (2009)]{carr09} Carretta,~E., Bragaglia,~A., Gratton,~R., D'Orazi,~V., \& Lucatello,~S., 2009, A\&A, 508, 695 

\bibitem[Catelan et~al.\ (2010)]{catelan10} Catelan,~M.,
Valcarce,~A.~A.~R., \& Sweigart,~A.~V., 2010, in IAU Symp.\ 266, Star
Clusters: Basic Galactic Building Blocks throughout Time and
Space, 281 


\bibitem[Clement et~al.\ (2001)]{clement2001} Clement,~C.~M., et~al.\ 2001, AJ, 122, 2587

\bibitem[Clementini et~al.\ (1994)]{clementini1994} Clementini,~G., Merighi,~R., Pasquini,~L., Cacciari,~C., \& Gouiffes,~C. 1994, MNRAS, 267, 83

\bibitem[Coppola et~al.\ (2011)]{coppola2011} Coppola,~G., et~al.\ 2011, MNRAS, 416, 1056


\bibitem[de~Sitter (1947)]{desitter47} de~Sitter,~A.\ 1947, BAN, 10, 287



\bibitem[Djorgovski et~al.\ (2011)]{djorg11} Djorgovski,~S.~G., et~al.\ 2011, ArXiv e-prints, arXiv:1102.5004

\bibitem[Dotter (2008)]{dotter2008} Dotter,~A., 2008, ApJ, 687, L21

\bibitem[Drake et~al.\ (2009)]{drake09} Drake,~A.~J., et~al.\ 2009, ApJ , 696, 870


\bibitem[Freedman et~al.\ (2012)]{freed12} Freedman,~W., et~al.\ 2012, Spitzer Proposal, 90002

Shimasaku,~K., \& Schneider,~D.~P. 1996, AJ, 111, 1748

\bibitem[Gezari et~al.\ (2013)]{geza13} Gezari,~S., et~al.\ 2013, ApJ , 766, 60

\bibitem[Goranskij (1980)]{gora80} Goranskij,~V.~P. 1980 ATsir, 1111, 6

\bibitem[Greenstein (1939)]{green39} Greenstein,~J.~L. 1939 ApJ 90, 401


\bibitem[Hansen et~al.\ (2004)]{hansen2004} Hansen,~B.~M.~S., et~al.\ 2004, ApJS, 155, 551

\bibitem[Harris (1996)]{harris96} Harris,~W.~E. 1996, AJ, 112,1487

\bibitem[Hendricks et~al.\ (2012)]{hendricks2012} Hendricks,~B. Stetson,~P.~B., VandenBerg,~D.~A., \& Dall'Ora,~M. 2012, AJ, 144, 25


\bibitem[Ivans et~al.\ (1999)]{ivans1999} Ivans,~I.~I., Sneden,~C., Kraft,~R.~P., Suntzeff,~N.~B., Smith,~V.~V., Langer,~G.~E., \& Fulbright,~J.~P. 1999, AJ, 118, 1273

\bibitem[Jurcksik et~al.\ (2009)]{jurc2009} Jurcsik,~J., et~al.\ 2009, MNRAS, 400, 1006

\bibitem[Kaluzny et~al.\ (1997)]{kaluz1997} Kaluzny,~J., Thompson,~I.~B., Krzeminski,~W., Pych,~W., Rucinski,~S.~M., Burley,~G.~S., \& Shectman,~S.~A. 1997, AJ, 113, 2219 

\bibitem[Kaluzny et~al.\ (2013a)]{kaluz2013a} Kaluzny,~J., Thompson,~I.~B. Rozyczka,~M., Dotter,~A., \& Krzeminski,~W. 2013, AJ, 145, 43 

\bibitem[Kaluzny et~al.\ (2013b)]{kaluz2013b} Kaluzny,~J., Thompson,~I.~B. Rozyczka,~M., \& Krzeminski,~W. 2013, AcA, 63, 181

\bibitem[Kinemuchi et~al.\ (2006)]{kine06} Kinemuchi,~K., Smith,~H.~A., Wo\`niak,~P.~R., McKay,~T.~A., \& ROTSE Collaboration. 2006, AJ , 132, 1202



\bibitem[Kunder et~al.\ (2011)]{kunder2011} Kunder,~A., et~al.\ 2011, AJ, 141, 15

\bibitem[Kunder et~al.\ (2013a)]{kunder2013a} Kunder,~A., Salaris,~M., Cassisi,~S., De~Propris,~R., Walker,~A., Stetson,~P.~B., Catelan,~M., \& Amigo,~P. 2013, AJ, 145, 25

\bibitem[Kunder et~al.\ (2013b)]{kunder2013b} Kunder,~A., et~al.\ 2013, AJ, 145, 33 

\bibitem[Kunder et~al.\ (2013c)]{kunder2013c} Kunder,~A., et~al.\ 2013, AJ, 146, 119 

\bibitem[Landolt (1973)]{landolt1973} Landolt,~A.~U. 1973, AJ, 78, 959

\bibitem[Landolt (1983)]{landolt1983} Landolt,~A.~U. 1983, AJ, 88, 439

\bibitem[Landolt (1992)]{landolt1992} Landolt,~A.~U. 1992, AJ, 104, 340

\bibitem[Layden (1998)]{lay98} Layden,~A.~C. 1998, AJ, 115, 193

\bibitem[Lee et~al.\ (1994)]{lee94} Lee,~Y.-W., Demarque,~P., \& Zinn,~R. 1994, ApJ, 423, 248

\bibitem[Libralato et~al.\ (2014)]{libralato2014} Libralato,~M., Bellini,~A., Bedin,~L.~R., Piotto,~G., Platais,~I., Kissler-Patig,~M., \& Milone,~A.~P. 2014, A\&A accepted, arXiv:1401.3344


\bibitem[Liu \& Janes (1990)]{liujanes1990} Liu,~T., \& Janes,~K.~A. 1990, ApJ, 360, 561


\bibitem[Lyne et~al.\ (1988)]{lyne1988} Lyne,~A.~G., Biggs,~J.~D., Brinklow,~A., McKenna,~J., \& Ashworth,~M. 1988 Nature, 332, 45




\bibitem[Malavolta et~al.\ (2014)]{malavolta2014} Malavolta,~L., Sneden,~C., Piotto,~G., Milone,~A.~P., Bedin,~L.~R., \& Nascimbeni,~V. 2014, AJ, 147, 25

\bibitem[Marchetti et~al.\ (2006)]{march2006} Marchetti, E. et~al.\ 2006, SPIE, 6262, 21

\bibitem[Marino et~al.\ (2008)]{marino2008} Marino,~A.~F., Villanova,~S., Piotto,~G., Milone,~A.~P., Momany,~Y., Bedin,~L.~R., \& Medling,~A.~M. 2008, A\&A, 490, 625

\bibitem[McCall (2004)]{mccall2004} McCall,~M.~L. 2004, AJ, 128, 2144

\bibitem[Miceli et~al.\ (2008)]{mice08} Miceli,~A., et~al.\ 2008, ApJ , 678, 865

\bibitem[Milone et~al.\ (2014)]{milone2014} Milone,~A.~P., et~al.\ 2014, MNRAS accepted, arXiv:1401.1091

\bibitem[Mochejska et~al.\ (2002)]{moche2002} Mochejska,~B.~J., Kaluzny,~J., \& Thompson,~I., Pych,~W. 2002, AJ, 124, 1486

\bibitem[Monelli et~al.\ (2013)]{monelli2013} Monelli,~M., et~al.\ 2013, MNRAS, 431, 2126

\bibitem[Mucciarelli et~al.\ (2011)]{mucciarelli2011} Mucciarelli,~A., Salaris,~M., Lovisi,~L., Ferraro,~F.~R., Lanzoni,~B., Lucatello,~S., \& Gratton,~R.~G. 2011, MNRAS, 412, 81


\bibitem[Palaversa et~al.\ (2013)]{pala13} Palaversa,~L., et~al.\ 2013, AJ , 146, 101


\bibitem[Pojmanski (2002)]{pojm2002} Pojmanski,~G. 2002, AcA , 52, 397

\bibitem[Pritzl et~al.\ (2004)]{pritzl2004} Pritzl,~B.~J., Venn,~K.~A., \& Irwin,~M. 2004, AJ, 130, 2140


\bibitem[Scargle (1982)]{scargle82} Scargle,~J.~D. 1982, ApJ, 263, 835

\bibitem[Schlegel et~al.\ (1998)]{schlegel1998} Schlegel,~D.~J., Finkbeiner,~D.~P., \& Davis,~M. 1998, ApJ, 500, 525

\bibitem[Sigudsson et~al.\ (2003)]{sigurd2003} Sigurdsson,~S., Richer,~H.~B., Hansen,~B.~M., Stairs,~I.~H., \& Thorsett,~S.~E. 2003, Science, 301, 193



\bibitem[S\'odor et~al.\ (2012)]{sodor2012} S\'odor,~\'A., et~al.\ 2012, MNRAS, 427, 1517

\bibitem[Sollima et~al.\ (2010)]{sollima2010} Sollima,~A., Cacciari,~C., Bellazzini,~M., \& Colucci,~S., 2010, MNRAS, 406, 329

\bibitem[Stellingwerf (2011)]{stellingw} Stellingwerf,~R.~F. 2011, RR~Lyrae Stars, Metal-Poor Stars, and the Galaxy, Carnegie Observatories Astrophysics Series, Vol. 5, 74

\bibitem[Stetson (1987)]{stetson1} Stetson,~P.~B. 1987, PASP, 99, 191

\bibitem[Stetson (1989)]{stetson89} Stetson,~P.~B. 1989, in ``Image and Data Processing; Interstellar Dust,'' V Escola Avan\c ada de Astrof\'isica, Departamento de Astronomia, Instituto Astron\^omico e Geof\'isico, Universidade de S\~ao Paulo, p. 1


\bibitem[Stetson (1996)]{stetson96} Stetson,~P.~B. 1996, PASP, 108, 851

\bibitem[Stetson (2000)]{stetson2000} Stetson,~P.~B. 2000, PASP, 112, 925

\bibitem[Stetson (2005)]{stetson2005} Stetson,~P.~B. 2005, PASP, 117, 563

\bibitem[Stetson et~al.\ (1998)]{stetson98} Stetson,~P.~B., et~al.\ 1998, ApJ, 508, 491


\bibitem[Sturch (1977)]{sturch1977} Sturch,~C.~R. 1977, PASP, 89, 349

\bibitem[Szabados \& Dora (2012)]{szab2012} Szabados,~L., \& Dora,~N. 2012, MNRAS, 426, 3148

\bibitem[Szeidl et~al.\ (2011)]{szeidl2011} Szeidl,~B., Hurta,~Z., Jurcsik,~J., Clement,~C., \& Lovas,~M. 2011, MNRAS, 411, 1744


\bibitem[VandenBerg et~al.\ (2013)]{vandenberg2013} VandenBerg,~D.~A., Broogard,~K., Leaman,~R., \& Casagrande,~L. 2013, ApJ, 775, 134

\bibitem[Vivas et~al.\ (2004)]{vivas04} Vivas,~A.~K., et~al.\ 2004, AJ , 127, 1158

\bibitem[Welch \& Stetson (1993)]{welchstet} Welch,~D.~L., \& Stetson,~P.~B. 1993, AJ, 105, 1813

\bibitem[Yao (1977)]{yao77} Yao,~B.-A. 1977, Acta Ast Sin, 18, 216

\bibitem[Yong et~al.\ (2008)]{yong2008} Yong,~D., Karakas,~A.~I., Lambert,~D.~L., Chieffi,~A., \& Limongi,~M. 2008, ApJ, 689, 1031

\end{thebibliography}

\clearpage
\begin{deluxetable}{lllll ccccc c}
\tablewidth{0pt}
\tabletypesize{\scriptsize}
\tablecaption{Log of observations in optical bands.}
\tablehead{ & Run ID & Dates & Telescope &  Camera & $U$ & $B$ & $V$ & $R$ & $I$ & Multiplex }
\startdata

 1 & pwm      &  1994 Apr 27         & JKT 1.0m      &  EEV7                &   --- &  --- &   5 & ---  & --- & --- \\ 
 2 & apr97    &  1997 Apr 14         & ESO  0.9m     &  ccd\$33             &   --- &  --- &   8 & ---  &  8 & --- \\ 
 3 & bond5    &  1997 Jun 02         & CTIO 0.9m     &  Tek2K$_3$           &    6 &   5 &   5 & ---  &  5 & --- \\ 
 4 & manu     &  2000 Jun 07--09     & ESO/Danish 1.5m&  DFOSC Loral 2kx2k  &    1 & 146 &   5 & ---  & --- & --- \\ 
 5 & wfi10    &  2000 Jul 08         & MPI/ESO 2.2m  &  WFI                 &   --- &   2 &   3 & ---  & --- &  8 \\ 
 6 & bond7    &  2001 Mar 28         & CTIO 0.9m     &  Tek2K$_3$           &    2 &   2 &   2 & ---  &  2 & --- \\ 
 7 & danish(a)&  2001 Apr 14--18     & Danish 1.5m   &  EEV 2kx4k           &   --- & 149 & 258 &  375& --- & --- \\ 
 8 & danish(b)&  2001 Apr 27--May 05 & Danish 1.5m   &  EEV 2kx4k           &   --- & 260 & 340 &  567& --- & --- \\ 
 9 & danish(c)&  2001 May 08--13     & Danish 1.5m   &  EEV 2kx4k           &   --- &  78 & 127 &  225& --- & --- \\ 
10 & danish(d)&  2001 May 25--30     & Danish 1.5m   &  EEV 2kx4k           &   --- & 103 & 171 &  304& --- & --- \\ 
11 & danish(e)&  2001 Jun 24--Jul 01 & Danish 1.5m   &  EEV 2kx4k           &   --- & 252 & 473 &  334& --- & --- \\ 
12 & not017   &  2001 Jul 10         & NOT 2.6m      &  CCD7                &   --- &   1 &   3 & ---  &  4 & --- \\ 
13 & wfi6     &  2002 Feb 21         & MPI/ESO 2.2m  &  WFI                 &   --- &   4 &   4 & ---  &  4 &  8 \\ 
14 & wfi5     &  2002 Jun 18-21      & MPI/ESO 2.2m  &  WFI                 &    4 &  12 &  14 & ---  & 72 &  8 \\ 
15 & fors20602&  2006 Feb 28         & ESO VLT 8.0m  &  FORS2 MIT/LL mosaic &   --- &  --- &   2 & ---  &  2 &  2 \\ 
16 & fors20605&  2006 May 29-30      & ESO VLT 8.0m  &  FORS2 MIT/LL mosaic &   --- &  --- &   1 &  4  &  1 &  2 \\ 
17 & emmi8    &  2007 Jul 13-16      & ESO NTT 3.6m  &  EMMI MIT/LL mosaic  &   --- &  19 &  19 & 11  & 15 &  2 \\ 
18 & Y1007    &  2010 Aug 03-04      & CTIO 1.0m     &  Y4KCam ITL SN3671   &   --- &  84 &  84 & ---  & --- & --- \\ 

\enddata

\tablecomments{\\
 1  Observer: ``PWM'' \\ 
 2  Observer:  A.~Rosenberg; data contributed by the observer \\ 
 3  Observer:  H.~E.~Bond; data contributed by the observer \\ 
 4  Observer:  unknown; data contributed by M.~Zoccali \\ 
 5  Observer:  Ferraro; ESO program identification:  065.L-0463 \\ 
 6  Observer:  H.~E.~Bond; data contributed by the observer \\ 
 7  Observer:  unknown \\ 
 8  Observer:  unknown \\ 
 9  Observer:  unknown \\ 
10  Observer:  unknown \\ 
11  Observer:  unknown \\ 
12  Observer:  unknown (probably Bruntt$?$) \\ 
13  Observer:  unknown; ESO program identification:  68.D-0265(A) \\ 
14  Observer:  unknown; ESO program identification:  69.D-0582(A) \\ 
15  Observer:  unknown; ESO program identification:  077.D-0775(B) \\ 
16  Observer:  unknown; ESO program identification:  077.D-0775(A) \\ 
17  Observer:  La-Silla-Sciops; ESO program identification:  59.A-9000 \\ 
18  Observer:  J.-W.~Lee; SMARTS Project ID:  Sejong10B}
\label{tab:table_obs_opt}
\end{deluxetable}

\begin{deluxetable}{lllll ccc c}
\tablewidth{0pt}
\tabletypesize{\scriptsize}
\tablecaption{Log of observations in NIR bands.}
\tablehead{ & Run ID & Dates & Telescope &  Camera & $J$ & $H$ & $K$ & Multiplex }
\startdata

 1 & m4ir     &  2002 Aug 16     &  ESO-NTT 3.5m &   SOFI      &     50 & --- &  50 & ---   \\
 2 & mad3     &  2007 Jun 04--06 &  ESO VLT 8.0m &   MAD       &    --- & --- & 152 & ---\\
 3 & hawki11  &  2007 Aug 03--05 &  ESO VLT 8.0m &   HAWKI     &     76 & --- &   5 &   4 \\
 4 & N10      &  2010 Sep 21     &  CTIO 4.0m    &   NEWFIRM   &     12 & --- & --- &   4 \\
 5 & hawki3   &  2011 Mar 27     &  ESO VLT 8.0m &   HAWKI     &     25 & --- & --- &   4 \\
 6 & hawki4   &  2011 Apr 26     &  ESO VLT 8.0m &   HAWKI     &     25 & --- & --- &   4 \\
 7 & N1105    &  2011 May 19     &  CTIO 4.0m    &   NEWFIRM   &     26 &  13 &  24 &   4 \\
 8 & hawki12  &  2011 Aug 23     &  ESO VLT 8.0m &   HAWKI     &     45 & --- & --- &   4 \\
 9 & hawki7   &  2012 May 16     &  ESO VLT 8.0m &   HAWKI     &     20 &   5 &  22 &   4 \\
10 & hawki5   &  2012 May 17     &  ESO VLT 8.0m &   HAWKI     &     16 & --- &  21 &   4 \\
11 & hawki8   &  2012 Jun 21     &  ESO VLT 8.0m &   HAWKI     &     62 &  12 &  21 &   4 \\
12 & hawki6   &  2012 Jun 22     &  ESO VLT 8.0m &   HAWKI     &    --- &  16 &  21 &   4 \\
13 & hawki10  &  2012 Jun 27--28 &  ESO VLT 8.0m &   HAWKI     &     53 &  43 &  42 &   4 \\
14 & hawki9   &  2012 Jul 14--15 &  ESO VLT 8.0m &   HAWKI     &     16 &  20 &  26 &   4 \\
\enddata

\tablecomments{\\
 1 ESO Program ID 69.D-0604(A) \\
 2 ESO Program ID Maintenance; PI-COI name Condor \\
 3 ESO Program ID 60.A-9283(A) \\
 4 NOAO Proposal ID 2010A-0036; observers Probst, Zeballos \\
 5 ESO Program ID 60.A-9800(L) \\
 6 ESO Program ID 60.A-9800(L) \\
 7 NOAO Proposal ID 2011A-0644; observer Alonso-Garcia \\
 8 ESO Program ID 60.A-9800(L) \\
 9 ESO Program ID 089.D-0291(A) \\
10 ESO Program ID 089.D-0291(A) \\
11 ESO Program ID 089.D-0291(A), 60.A-9800(L) \\
12 ESO Program ID 089.D-0291(A) \\
13 ESO Program ID 089.D-0291(A) \\
14 ESO Program ID 089.D-0291(A)}
\label{tab:table_obs_ir}
\end{deluxetable}

\begin{deluxetable}{l c c cccc}
\tablewidth{0pt}
\tabletypesize{\footnotesize}
\tablecaption{Positions and periods for the RR Lyrae variables in M4.}
\tablehead{ID & 
\multicolumn{1}{c}{$\alpha$ (J2000.0)}   & 
\multicolumn{1}{c}{$\delta$ (J2000.0)}   & 
\multicolumn{3}{c}{Period\tablenotemark{a}}& 
$T_0$\tablenotemark{b} \\
        & h \quad m \quad s  & \deg \quad \farcm \quad \farcs & Clement & SL & LS & }
\startdata
V1  & 16 23 13.67 & --26 30 53.6 & 0.2889 & 0.28888237 & 0.28888260  &50601.4550 \\
V2  & 16 23 16.30 & --26 34 46.6 & 0.5357 & 0.53568190 & 0.53568192  &50601.2750 \\
V3  & 16 23 19.47 & --26 39 57.0 & 0.5066 &  ---    & 0.50667785  &  ---   \\
V5  & 16 23 21.03 & --26 33 05.8 & 0.6224 & 0.62240130 & 0.62240109  &50601.4778 \\
V6  & 16 23 25.79 & --26 26 16.3 & 0.3205 & 0.32051437 & 0.32051511  &49469.6457 \\
V7  & 16 23 25.95 & --26 27 41.9 & 0.4988 & 0.49878480 & 0.49878722  &50601.4511 \\
V8  & 16 23 26.16 & --26 29 41.6 & 0.5082 & 0.50822373 & 0.50822362  &50601.6924 \\
V9  & 16 23 26.80 & --26 29 48.0 & 0.5719 & 0.57189461 & 0.57189448  &50601.4580 \\
V10 & 16 23 29.21 & --26 28 54.3 & 0.4907 & 0.49071584 & 0.49071753  &50601.2890 \\
V11 & 16 23 29.98 & --26 36 24.5 & 0.4931 & 0.49321080 & 0.49320868  &52088.7604 \\
V12 & 16 23 30.82 & --26 34 58.9 & 0.4461 & 0.44610899 & 0.44610979  &50601.5210 \\
V14 & 16 23 31.29 & --26 35 34.5 & 0.4635 & 0.46353167 & 0.46353112  &50601.6818 \\
V15 & 16 23 31.96 & --26 24 18.1 & 0.4437 & 0.44371683 & 0.44366078  &50601.6418 \\
V16 & 16 23 32.50 & --26 30 23.0 & 0.5425 & 0.54254940 & 0.54254826  &50601.5688 \\
V18 & 16 23 34.70 & --26 31 04.6 & 0.4788 & 0.47879359 & 0.47879201  &50601.5170 \\
V19 & 16 23 35.05 & --26 25 36.3 & 0.4678 & 0.46781097 & 0.46781107  &50601.3741 \\
V20 & 16 23 35.39 & --26 32 35.9 & 0.3094 & 0.30941795 & 0.30941947  &50601.6644 \\
V21 & 16 23 35.93 & --26 31 33.6 & 0.4720 & 0.47200727 & 0.47200741  &50601.4735 \\
V22 & 16 23 36.95 & --26 30 13.0 & 0.6031 & 0.60306110 & 0.60306357  &50601.4997 \\
V23 & 16 23 37.33 & --26 31 56.1 & 0.2986 & 0.29861522 & 0.29861557  &50601.7160 \\
V24 & 16 23 38.04 & --26 30 41.8 & 0.5468 & 0.54678280 & 0.54678330  &50601.3191 \\
V25 & 16 23 39.42 & --26 30 21.3 & 0.6127 & 0.61273466 & 0.61273480  &50601.6449 \\
V26 & 16 23 41.65 & --26 32 41.1 & 0.5412 & 0.54121703 & 0.54121738  &50552.3694 \\
V27 & 16 23 43.17 & --26 27 16.3 & 0.6120 & 0.61201625 & 0.61201831  &50601.6926 \\
V28 & 16 23 53.60 & --26 30 05.3 & 0.5223 & 0.52234419 & 0.52234106  &50552.6908 \\
V29 & 16 23 58.25 & --26 21 35.1 & 0.5225 & 0.51766400 & 0.52248466  &46585.1061   \\
V30 & 16 23 59.65 & --26 32 34.3 & 0.2697 & 0.26974860 & 0.26974906  &50601.7050 \\
V31 & 16 24 00.64 & --26 30 42.0 & 0.5052 & 0.50520360 & 0.50520423  &50601.6277 \\
V32 & 16 24 29.70 & --26 32 02.5 & 0.5791 & 0.56578120 & 0.57910475  &42890.5531   \\
V33 & 16 24 33.33 & --26 20 58.1 & 0.6148 & 0.61483030 & 0.61483542  &43682.8820 \\
V34 & 16 22 33.64 & --26 24 46.6 & 0.5548 &   ---      &   ---       &29723.338  \\
V35 & 16 23 06.59 & --26 30 27.0 & 0.6270 & 0.62702470 & 0.62702374  &50601.4263 \\
V36 & 16 23 19.45 & --26 35 49.0 & 0.5413 & 0.54130500 & 0.54130918  &50601.2199 \\
V37 & 16 23 31.60 & --26 31 30.6 & 0.2474 & 0.24736150 & 0.24734352  &50601.5133 \\
V38 & 16 23 32.87 & --26 33 03.5 & 0.5778 & 0.57784690 & 0.57784632  &50601.3413 \\
V39 & 16 23 34.67 & --26 32 52.1 & 0.6240 & 0.62395110 & 0.62395399  &50601.4729 \\
V40 & 16 23 36.49 & --26 30 44.2 & 0.4015 & 0.38533000 & 0.38533005  &50601.6088 \\
V41 & 16 23 39.50 & --26 33 59.8 & 0.2517 & 0.25174170 & 0.25174181  &50601.5357 \\
V42 & 16 24 02.00 & --26 22 13.0 & 0.3037 & 0.30685490 & 0.31733120  &46585.4073 \\
V43\tablenotemark{c} & 16 25 07.47 & --26 25 43.9 & 0.3206 & 0.32066000 & 0.32068160  &43681.1370\\
V49 & 16 23 45.25 & --26 31 28.4 & 0.2160 & 0.22754460 & 0.22754331  &50552.2894 \\
V52 & 16 23 24.06 & --26 30 27.8 & 0.4605 & 0.85549650 & 0.85549784  &50601.0100 \\
V61 & 16 23 29.76 & --26 29 50.3 &  ---   & 0.26528682 & 0.26528645  &50601.5119 \\
V64 & 16 23 15.06 & --26 30 19.1 & 0.6033 & 0.60448250 & 0.60448407  &50601.1695 \\
V76 & 16 22 05.56 & --26 21 43.1 & 0.3058 & --- & --- & ---  \\
C1  & 16 23 36.43 & --26 30 42.8 &  ---   & 0.28625730 & 0.28625730  &50601.4604 \\
C2  & 16 24 07.59 & --26 38 38.2 &  ---   & 0.45480186 & 0.45480258  &50601.6896 \\

\enddata
\tablenotetext{a}{Pulsation periods (days) based on a compilation of literature values \citep{clement2001} or 
on our current data using either the string-length method (SL) or the 
Lomb--Scargle method (LS).} 
\tablenotetext{b}{Epoch of maximum (HJD--2,400,000) estimated from
the analytical fit of the $V$-band light curve, except V33 (T0
from \citet{cacc1979}) and V34 from \citet{desitter47}} 
\tablenotetext{c}{We have no data of our own for this star.  The
astrometry is based on measurements of three images from the
Digitized Sky Survey, carefully transformed to our coordinate
system.  This is the position of a fairly isolated star of
appropriate apparent brightness lying 0\Sec7 from the position
given in \cite{clement2001}.  The light-curve parameters are based
on data from \citet{cacc1979}.}
\label{tab:pos_rr}
\end{deluxetable}

\begin{deluxetable}{ll ccccc ccccc l}
\tablewidth{0pt}
\tabletypesize{\footnotesize}
\tablecaption{Mean optical magnitudes and amplitudes for the RR Lyraes in M4.}
\tablehead{
ID & 
Period\tablenotemark{a}& 
$U$\tablenotemark{b}& 
$B$\tablenotemark{b}& 
$V$\tablenotemark{b}& 
$R$\tablenotemark{b}& 
$I$\tablenotemark{b}& 
$A_U$\tablenotemark{c} & 
$A_B$\tablenotemark{c} & 
$A_V$\tablenotemark{c} & 
$A_R$\tablenotemark{c} & 
$A_I$\tablenotemark{c} & 
Mode\tablenotemark{d} \\
                       & days & mag & mag  & mag  & mag & mag & mag & mag & mag & mag  & mag & }
\startdata

V1   & 0.28888260 & 14.56 &  14.02 &  13.43 &  13.00 &  12.54 & 0.44 & 0.57 &   0.44 &   0.36 &   0.24 & RRc  \\
V2   & 0.53568192 & 14.50 &  14.12 &  13.41 &  12.91 &  12.39 & 1.09 & 1.26 &   0.96 &   0.76 &   0.61 & RRab*  \\
V3   & 0.50667785 & 14.71 &  13.65 &  13.01 &  ---   &  12.28 & ---  & 0.99 &  ---   &  ---   &  ---   & RRab  \\
V5   & 0.62240109 & 14.57 &  14.12 &  13.37 &  12.84 &  12.28 & ---  & 0.43 &   0.33 &   0.26 &   0.20 & RRab  \\
V6   & 0.32051511 & 14.41 &  14.08 &  13.45 &  13.00 &  12.50 & 0.63 & 0.56 &   0.43 &   0.35 &   0.25 & RRc  \\
V7   & 0.49878722 & 14.56 &  14.09 &  13.42 &  12.94 &  12.40 & ---  & 1.39 &   1.06 &   0.86 &   0.70 & RRab  \\
V8   & 0.50822362 & 13.96 &  13.98 &  13.32 &  12.85 &  12.33 & ---  & 1.41 &   1.11 &   0.92 &   0.69 & RRab  \\
V9   & 0.57189448 & 14.48 &  13.98 &  13.30 &  12.82 &  12.30 & ---  & 1.44 &   1.11 &   0.91 &   0.71 & RRab  \\
V10  & 0.49071753 & 14.66 &  13.95 &  13.33 &  12.88 &  12.38 & ---  & 1.60 &   1.25 &   1.02 &   0.79 & RRab  \\
V11  & 0.49320868 & 14.53 &  14.07 &  13.40 &  12.93 &  12.41 & 0.92 & 1.24 &   1.07 &   0.86 &   0.65 & RRab*  \\
V12  & 0.44610979 & 14.49 &  14.23 &  13.58 &  13.10 &  12.54 & 1.60 & 1.61 &   1.28 &   1.06 &   0.81 & RRab  \\
V14  & 0.46353112 & 14.63 &  14.26 &  13.59 &  13.10 &  12.58 & 1.63 & 1.60 &   1.23 &   1.02 &   0.79 & RRab* \\
V15  & 0.44366078 & 14.99 &  14.45 &  13.69 &  13.08 &  12.63 & ---  & 1.18 &   0.92 &   0.78 &   0.53 & RRab*  \\
V16  & 0.54254826 & 14.47 &  14.05 &  13.34 &  12.85 &  12.31 & ---  & 1.15 &   0.89 &   0.72 &   0.57 & RRab  \\
V18  & 0.47879201 & 14.66 &  14.02 &  13.36 &  12.88 &  12.35 & ---  & 1.45 &   1.12 &   0.92 &   0.71 & RRab  \\
V19  & 0.46781107 & 14.26 &  14.01 &  13.38 &  12.92 &  12.38 & 1.46 & 1.58 &   1.24 &   1.02 &   0.78 & RRab  \\
V20  & 0.30941947 & 14.27 &  13.72 &  13.19 &  12.80 &  12.33 & ---  & 0.36 &   0.23 &   0.22 &   0.14 & RRc  \\
V21  & 0.47200741 & 14.40 &  13.88 &  13.19 &  12.70 &  12.16 & ---  & 1.43 &   1.13 &   0.85 &   0.68 & RRab  \\
V22  & 0.60306357 & 14.60 &  14.08 &  13.33 &  12.81 &  12.24 & ---  & 0.66 &   0.48 &   0.40 &   0.30 & RRab*  \\
V23  & 0.29861557 & 14.28 &  13.75 &  13.19 &  12.78 &  12.41 & ---  & 0.56 &   0.43 &   0.35 &   0.27 & RRc  \\
V24  & 0.54678330 & 14.60 &  14.02 &  13.33 &  12.84 &  12.31 & ---  & 1.08 &   0.86 &   0.71 &   0.53 & RRab*  \\
V25  & 0.61273480 & 14.07 &  13.94 &  13.23 &  12.74 &  12.20 & ---  & 1.09 &   0.83 &   0.69 &   0.38 & RRab  \\
V26  & 0.54121738 & 14.46 &  13.87 &  13.25 &  12.79 &  12.27 & ---  & 1.57 &   1.24 &   1.01 &   0.78 & RRab  \\
V27  & 0.61201831 & 14.22 &  13.90 &  13.21 &  12.73 &  12.18 & 1.25 & 1.19 &   0.91 &   0.73 &   0.56 & RRab  \\
V28  & 0.52234106 & 14.07 &  13.79 &  13.19 &  12.75 &  12.23 & 1.16 & 1.27 &   1.01 &   0.78 &   0.90 & RRab*  \\
V29  & 0.52248466 & 14.20 &  13.90 &  13.28 &  12.83 &  12.31 & 1.10 & 1.40 &   1.08 &   0.96 &   0.79 & RRab* \\
V30  & 0.26974906 & 14.16 &  13.84 &  13.32 &  12.94 &  12.48 & 0.58 & 0.66 &   0.51 &   0.40 &   0.32 & RRc  \\
V31  & 0.50520423 & 14.12 &  13.77 &  13.20 &  12.77 &  12.32 & 1.42 & 1.13 &   0.86 &   0.67 &   0.55 & RRab*  \\
V32  & 0.57910475 & 14.12 &  13.83 &  13.18 &  12.70 &  12.04 & 0.97 & 1.05 &   0.76 &   0.64 &   0.51 & RRab  \\
V33  & 0.61483542 & 13.94 &  13.65 &  13.03 &  12.58 &  12.08 & 1.23 & 1.21 &   0.95 &  ---   &   0.59 & RRab  \\
V34  & 0.5548\tablenotemark{e} & 15.47 &  14.73 &  14.02 &  ---   &  12.77 & ---  & ---&  ---   &  ---   &  ---   & RRab  \\
V35  & 0.62702374 & 14.42 &  14.14 &  13.38 &  12.84 &  12.30 & 0.75 & 0.87 &   0.66 &   0.51 &   0.42 & RRab* \\
V36  & 0.54130918 & 14.42 &  14.13 &  13.42 &  13.03 &  12.40 & ---  & 1.20 &   0.92 &   0.73 &   0.59 & RRab  \\
V37  & 0.24734352 & 14.45 &  13.90 &  13.37 &  12.99 &  12.51 & ---  & 0.28 &   0.22 &   0.18 &   0.14 & RRc  \\
V38  & 0.57784632 & 14.59 &  14.20 &  13.42 &  12.87 &  12.29 & ---  & 0.85 &   0.63 &   0.49 &   0.32 & RRab*  \\
V39  & 0.62395399 & 14.66 &  14.20 &  13.43 &  12.89 &  12.29 & ---  & 0.55 &   0.42 &   0.34 &   0.27 & RRab*  \\
V40  & 0.38533005 & 14.27 &  13.74 &  13.13 &  12.70 &  12.20 & ---  & 0.56 &   0.43 &   0.34 &   0.27 & RRc  \\
V41  & 0.25174181 & 14.61 &  14.03 &  13.49 &  13.09 &  12.64 & ---  & 0.45 &   0.36 &   0.28 &   0.25 & RRc  \\
V42  & 0.30685490 & 14.14 &  13.78 &  13.24 &  12.87 &  12.36 & ---  & 0.54 &   0.41 &   0.33 &   0.27 & RRc  \\
V43\tablenotemark{f} & 0.32066000 & 13.86 &  13.60 &  13.08 &  ---  &  ---  &  0.52 &  0.58 &   0.43 &  ---   &   ---& RRc  \\
V49  & 0.22754331 & 14.39 &  13.81 &  13.32 &  12.97 &  12.55 & ---  & 0.13 &   0.08 &   0.07 &   0.05 & RRc  \\
V52  & 0.85549784 & 14.28 &  13.90 &  13.11 &  12.56 &  11.97 & 0.41 & 0.39 &   0.31 &   0.25 &   0.19 & RRab  \\
V61  & 0.26528645 & 14.35 &  13.78 &  13.26 &  12.90 &  12.46 & ---  & 0.21 &   0.17 &   0.14 &   0.09 & RRc  \\
V64  & 0.60448407 & 21.42 &  20.74 &  20.01 &  19.50 &  18.93 & ---  & 1.08 &   0.84 &   0.68 &   0.54 & RRab*  \\
V76\tablenotemark{e} & 0.3058 & 15.44 & 14.32 & 13.68 & --- & 12.62 & --- & --- & --- & --- & --- & RR \\
C1   & 0.28625730 & 14.37 &  13.89 &  13.32 &  12.92 &  12.45 & 0.60 & 0.59 &   0.46 &   0.38 &   0.25 & RRc  \\
C2   & 0.45480258 & 17.46 &  16.71 &  16.04 &  15.83 &  15.07 & ---  & 1.63 &   1.23 &  ---   &   0.77 & RRab  \\
\enddata
\tablenotetext{a}{Pulsation period based on the LS method.}
\tablenotetext{b}{Mean $U,B,V,R,I$ magnitudes. The mean was computed as an intensity mean 
and then transformed into magnitude.} 
\tablenotetext{c}{{L}uminosity amplitudes in $U,B,V,R,I$. The amplitudes were estimated as 
the difference netween minimum and maximum of the analytical fit. The $U$ amplitude is 
only available for a few variables (see Appendix).} 
\tablenotetext{d}{Pulsation mode: RRab, fundamental; RRc, first overtone. }
\tablenotetext{e}{The period is the one given in the compilation
of Clement.  The tabulated magnitudes based on robust means of our
data; we have insufficient data to fit a light curve.} 
\tablenotetext{f}{Tabulated results based on published data from \citet{cacc1979}; the star lies outside our field.}
\label{tab:phot_rr_opt}
\end{deluxetable}

\begin{deluxetable}{ll ccc ccc l}
\tablewidth{0pt}
\tabletypesize{\footnotesize}
\tablecaption{Mean NIR magnitudes and amplitudes for the RR Lyraes in M4.}
\tablehead{
ID & 
Period\tablenotemark{a}& 
$J$\tablenotemark{b}& 
$H$\tablenotemark{b}& 
$K$\tablenotemark{b}& 
$A_J$\tablenotemark{c} & 
$A_H$\tablenotemark{c} & 
$A_K$\tablenotemark{c} & 
Mode\tablenotemark{d} \\
                      & days & mag & mag  & mag  & mag & mag & mag & }
\startdata

V1   & 0.28888260 & 11.79 & 11.50 & 11.42 &  0.12 &  0.09 &  0.08 &  RRc  \\
V2   & 0.53568192 & 11.57 & 11.10 & 11.07 &  0.38 & --- &  0.28 &  RRab*  \\
V3   & 0.50667785 & 11.39 & 11.27 & 11.07 &  0.43 & --- & ---&  RRab  \\
V5   & 0.62240109 & 11.40 & 11.17 & 10.94 &  0.19 & --- & ---&  RRab  \\
V6   & 0.32051511 & 11.69 & 11.45 & 11.31 &  0.18 & --- & ---&  RRc  \\
V7   & 0.49878722 & 11.59 & 11.16 & 11.14 & ---& --- & ---&  RRab  \\
V8   & 0.50822362 & 11.57 & 11.40 & 11.15 &  0.36 & --- & ---&  RRab  \\
V9   & 0.57189448 & 11.45 & 11.02 & 10.96 &  0.54 & --- & ---&  RRab  \\
V10  & 0.49071753 & 11.52 & 11.35 & 11.20 &  0.33 & --- & ---&  RRab  \\
V11  & 0.49320868 & 11.64 & 11.22 & 11.22 &  0.36 & --- & ---&  RRab*  \\
V12  & 0.44610979 & 11.73 & 11.36 & 11.25 &  0.48 & --- & ---&  RRab  \\
V14  & 0.46353112 & 11.58 & 11.30 & 11.25 &  0.30  & --- & ---&  RRab* \\
V15  & 0.44366078 & 11.74 & 11.56 & 11.25 &  0.37 & --- &  0.28 &  RRab*  \\
V16  & 0.54254826 & 11.49 & 11.14 & 10.99 &  0.32 & --- & ---&  RRab  \\
V18  & 0.47879201 & 11.63 & 11.40 & 11.08 &  0.42 & --- & ---&  RRab  \\
V19  & 0.46781107 & 11.60 & 11.23 & 11.18 &  0.46 & --- & ---&  RRab  \\
V20  & 0.30941947 & 11.58 & 11.41 & 11.26 &  0.18 & --- & ---&  RRc  \\
V21  & 0.47200741 & 11.59 & 11.22 & 11.14 &  0.41 & --- & ---&  RRab  \\
V22  & 0.60306357 & 11.38 & 11.03 & 10.89 &  0.31  & --- & ---&  RRab*  \\
V23  & 0.29861557 & 11.65 & 11.40 & 11.29 &  0.08 & --- & ---&  RRc  \\
V24  & 0.54678330 & 11.47 & 11.13 & 11.12 &  0.37 & --- & ---&  RRab*  \\
V25  & 0.61273480 & 11.35 & 11.15 & 10.90 &  0.40 & --- & ---&  RRab  \\
V26  & 0.54121738 & 11.51 & 11.24 & 11.02 &  0.46 & --- & ---&  RRab  \\
V27  & 0.61201831 & 11.41 & 10.95 & 10.90 &  0.39 & --- & ---&  RRab  \\
V28  & 0.52234106 & 11.53 & 11.15 & 11.07 &  0.52 & --- & ---&  RRab*  \\
V29  & 0.52248466 & 11.44 & 11.16 & 11.04 &  0.31  & --- & ---&  RRab* \\
V30  & 0.26974906 & 11.82 & 11.59 & 11.48 &  0.16 & --- & ---&  RRc  \\
V31  & 0.50520423 & 11.35 & 11.15 & 11.13 &  0.44 & --- & ---&  RRab*  \\
V32  & 0.57910475 & 11.42 & 11.04 & 10.93 &  0.31 & --- &  0.17&  RRab  \\
V33  & 0.61483542 & 11.35 & 10.92 & 10.88 &  0.41 & --- &  0.29&  RRab  \\
V34  & 0.5548\tablenotemark{e} & 11.50 & 11.21 & 11.06 & ---& --- & ---&  RRab  \\
V35  & 0.62702374 & 11.41 & 11.06 & 10.91 &  0.17 &  0.25 &  0.24&  RRab* \\
V36  & 0.54130918 & 11.57 & 11.21 & 11.08 &  0.28 & --- & ---&  RRab  \\
V37  & 0.24734352 & 11.84 & 11.69 & 11.51 &  0.06 &  0.07 & ---&  RRc  \\
V38  & 0.57784632 & 11.36 & 11.10 & 10.89 &  0.15 & --- & ---&  RRab*  \\
V39  & 0.62395399 & 11.41 & 11.02 & 10.91 &  0.23 & --- & ---&  RRab*  \\
V40  & 0.38533005 & 11.42 & 11.22 & 11.04 &  0.13 & --- & ---&  RRc  \\
V41  & 0.25174181 & 11.90 & 11.66 & 11.58 &  0.14 & --- & ---&  RRc  \\
V42  & 0.30685490 & 11.74 & 11.45 & 11.35 &  0.18 & --- & ---&  RRc  \\
V43\tablenotemark{f}  & 0.32066000 & ---   & ---   & ---   & ---& --- & ---&  RRc  \\
V49  & 0.22754331 & 11.89 & 11.72 & 11.60 &  0.06 & --- & ---&  RRc  \\
V52  & 0.85549784 & 11.06 & 10.68 & 10.60 &  0.13 & --- & ---&  RRab  \\
V61  & 0.26528645 & 11.77 & 11.55 & 11.43 &  0.11 & --- & ---&  RRc  \\
V64  & 0.60448407              & 18.05 &   --- & 17.40 &  0.16 & --- & ---&  RRab*  \\
V76\tablenotemark{e} & 0.3058 & 15.44 & 14.32 & 13.68 &   --- & --- & ---& RR \\
C1   & 0.28625730              & 11.69 & 11.46 & 11.35 &  0.17 & --- & ---&  RRc  \\
C2   & 0.45480258 & 14.28 & 13.76 & 13.72 &  0.2$\,\,\,$ & --- & ---&  RRab  \\
\enddata
\tablenotetext{a}{Pulsation period based on the LS method.} 
\tablenotetext{b}{Mean J,H,K magnitudes. The mean was computed as an intensity mean 
and then transformed into magnitude.} 
\tablenotetext{c}{{L}uminosity amplitudes in J,H,K. The amplitudes were estimated as 
the difference netween minimum and maximum of the analytical fit.} 
\tablenotetext{d}{Pulsation mode: $RR_{ab}$, fundamental; $RR_{c}$, first overtone. 
An asterisk marks the variables that are candidate Blazhko variables.} 
\tablenotetext{e}{The period is the one given in the compilation
of Clement.  The tabulated magnitudes based on robust means of our
data; we have insufficient data to fit a light curve.} 
\tablenotetext{f}{Outside our field of view.}
\label{tab:phot_rr_ir}
\end{deluxetable}

\begin{deluxetable}{l cc ccccc l }
\tablewidth{0pt}
\tabletypesize{\footnotesize}
\tablecaption{Positions and mean optical photometry for other candidate variable stars.}
\tablehead{ID & 
\multicolumn{1}{c}{$\alpha$ (J2000.0)} & 
\multicolumn{1}{c}{$\delta$ (J2000.0)} &
$U$ &
$B$ &
$V$ &
$R$ &
$I$ &
period (d) or \\
              & h \quad m \quad s  & \deg\quad \farcm \quad \farcs & & & & & & comment}
\startdata
V53     & 16 23 14.43 & --26 36 05.4 & 14.64 & 12.73 & 10.88 &  9.80 &  9.04 & variable? \\
V54     & 16 23 42.82 & --26 29 27.5 & 17.79 & 13.98 & 12.66 & 11.86 & 11.06 & LPV or EB? \\
V55     & 16 23 45.93 & --26 23 36.9 & 14.22 & 13.70 & 13.24 &  ---  & 12.59 & LPV or EB? \\
V56     & 16 23 45.95 & --26 33 38.8 & 15.02 & 14.24 & 12.93 & 12.13 & 11.35 & LPV or EB? \\
V57     & 16 23 21.17 & --26 31 59.7 & 14.58 & 14.12 & 12.99 & 12.28 & 11.56 & LPV or EB? \\
V58     & 16 23 47.84 & --26 32 05.9 & 14.30 & 13.83 & 13.50 & 13.24 & 12.94 & LPV or EB? \\
V59     & 16 23 50.15 & --26 33 23.8 & 14.66 & 14.35 & 13.44 & 12.81 & 12.17 & LPV or EB? \\
V60     & 16 23 45.23 & --26 33 56.8 & 15.54 & 14.82 & 13.56 & 12.79 & 12.02 & LPV or EB? \\
V63=K44 & 16 23 21.05 & --26 33 25.0 & 18.97 & 18.82 & 17.80 & 17.14 & 16.49 & 0.26358410 \\
V65=K46 & 16 23 47.19 & --26 31 56.5 & 18.85 & 18.85 & 18.58 & 18.39 & 18.16 & LPV or EB? \\
V66=K47 & 16 23 25.57 & --26 29 11.7 & 17.98 & 17.77 & 16.93 & 16.37 & 15.77 & 0.26987521 \\
V67=K48 & 16 23 36.81 & --26 31 44.1 & 17.27 & 17.12 & 16.25 & 15.65 & 15.09 & 0.28269412 \\
V68=K49 & 16 23 34.35 & --26 32 01.8 & 19.32 & 18.26 & 17.03 & 16.26 & 15.49 & 0.29744408 \\
V69=K50 & 16 23 31.34 & --26 31 48.4 & 18.27 & 18.18 & 17.32 & 16.74 & 16.21 & 0.26600006 \\
V70=K51 & 16 23 33.22 & --26 31 09.1 & 17.92 & 17.89 & 17.00 & 16.40 & 15.83 & 0.30368223 \\
V71=K52 & 16 23 31.50 & --26 30 57.7 & 17.84 & 17.67 & 16.70 & 16.03 & 15.39 & 0.776468 \\
V72=K53 & 16 23 38.51 & --26 32 10.8 & 16.76 & 16.46 & 15.83 & 15.40 & 15.01 & 0.30844869 \\
V73=K54 & 16 23 50.96 & --26 34 42.1 & 18.90 & 18.72 & 17.78 & 17.14 & 16.54 & 0.25246449 \\
V74=K55 & 16 23 45.78 & --26 31 16.5 & 17.64 & 17.70 & 16.87 & 16.31 & 15.72 & 0.31070264 \\
V75     & 16 22 51.68 & --26 25 14.7 & 14.50 & 14.31 & 13.37 &  ---  & 12.14 & insufficient data \\
V77=K56 & 16 23 34.30 & --26 29 55.9 & 16.32 & 15.83 & 14.66 & 13.92 & 23.21 & possible EB \\
V78=K57 & 16 23 36.73 & --26 31 50.6 & 16.57 & 16.24 & 15.20 & 14.53 & 13.86 & not variable? \\
V79     & 16 24 01.20 & --26 21 56.2 & 16.74 & 16.18 & 15.07 &  ---  & 13.69 & insufficient data \\
V80     & 16 23 48.94 & --26 23 51.9 & 17.05 & 16.66 & 15.60 &  ---  & 14.25 & not variable? \\
K59     & 16 23 39.18 & --26 29 54.8 &  ---  & 21.69 & 20.27 & 19.27 & 18.40 & 0.7114224 \\
K60     & 16 23 47.50 & --26 32 12.6 & 17.75 & 17.62 & 16.77 & 16.19 & 15.60 & 0.3703839 \\
K61     & 16 23 42.35 & --26 33 17.9 & 16.60 & 16.29 & 15.70 & 15.29 & 14.83 & not variable? \\
K62     & 16 23 32.62 & --26 36 13.1 & 20.05 & 19.74 & 19.13 & 18.72 & 18.22 & 0.04088083 \\
K63     & 16 23 27.37 & --26 32 26.6 & 16.80 & 17.78 & 17.67 & 17.55 & 17.42 & LPV or EB?\\
K64     & 16 23 20.80 & --26 31 02.2 & 19.43 & 19.07 & 18.44 & 18.02 & 17.55 & not variable? \\
K65     & 16 23 28.41 & --26 30 21.8 & 17.57 & 17.44 & 16.52 & 15.90 & 15.27 & 2.293172 \\
K66     & 16 23 32.26 & --26 31 41.1 & 17.71 & 17.64 & 16.78 & 16.20 & 15.64 & 8.11289 \\
K68     & 16 23 38.59 & --26 30 30.9 & 16.14 & 15.84 & 15.24 & 14.82 & 14.38 & not variable? \\
K69     & 16 23 58.11 & --26 37 18.9 & 17.93 & 17.89 & 17.01 & 16.43 & 15.81 & 48.195\tablenotemark{a} \\
C3      & 16 23 35.59 & --26 27 08.1 & 18.53 & 18.01 & 16.80 & 16.01 & 15.24 & 19.4651 \\
C4      & 16 23 44.79 & --26 24 29.3 & 16.30 & 16.08 & 15.17 & 14.58 & 13.98 & 0.4385898 \\
C5      & 16 23 34.60 & --26 25 41.4 & 20.39 & 19.82 & 19.00 & 18.45 & 18.02 & 0.41970602 \\
C6      & 16 23 59.77 & --26 29 43.6 & 19.32 & 19.12 & 18.06 & 17.37 & 16.72 & 0.26297749 \\
C7      & 16 24 02.14 & --26 30 51.9 & 21.32 & 20.15 & 18.90 & 18.13 & 17.28 & 0.27161155 \\
\enddata
\tablenotetext{a}{Because of the haphazard cadence of our observations, this star is subject
to extreme aliasing problems.  Left to ourselves, we would not have been able to determine a
secure period.  \cite{kaluz2013a} give the period of this star as 48.19$\,$d, and this is
the best period we were able to find between 48.1 and 48.3$\,$d.  Many other periods outside
this range would be possible if our data were all that were available.}
\label{tab:table_candopt}
\end{deluxetable}

\begin{deluxetable}{l rrr l }
\tablewidth{0pt}
\tabletypesize{\footnotesize}
\tablecaption{Mean infrared photometry for other candidate variable stars.}
\tablehead{ID & 
$J$ &
$H$ &
$K$ &
comment }
\startdata
V53     & 11.14 &  9.40 &  9.28 & probably saturated \\
V54     &  9.82 &  9.29 &  8.96 & probably saturated \\
V55     & 12.06 & 11.91 & 11.79 \\
V56     & 10.11 &  9.62 &  9.32 & probably saturated \\
V57     & 10.44 & 10.18 &  9.90 & probably saturated \\
V58     & 12.46 & 12.38 & 12.28 \\
V59     & 11.21 & 10.80 & 10.62 & possibly saturated \\
V60     & 10.80 & 10.37 &  9.98 & probably saturated \\
V63=K44 & 15.55 & 14.81 & 14.69 \\
V65=K46 & 17.82 & 17.44 & 17.06 \\
V66=K47 & 14.92 & 14.43 & 14.34 \\
V67=K48 & 14.74 & 14.50 & 14.47 \\
V68=K49 & 14.49 & 13.89 & 13.69 \\
V69=K50 & 15.38 & 14.98 & 14.58 \\
V70=K51 & 15.05 & 14.58 & 14.33 \\
V71=K52 & 14.57 & 13.99 & 13.79 \\
V72=K53 & 14.29 & 14.13 & 13.90 \\
V73=K54 & 15.57 & 14.98 & 14.75 \\
V74=K55 & 14.80 & 14.48 & 14.26 \\
V75     & 11.31 & 10.85 & 10.82 & possibly saturated \\
V77=K56 & 12.05 & 11.38 & 11.20 \\
V78=K57 & 12.78 & 12.22 & 12.06 \\
V79     & 12.68 & 12.08 & 11.87 \\
V80     & 13.25 & 12.63 & 12.48 \\
K58     & 15.00 & 14.52 & 14.37 \\
K59     & 16.97 & 16.00 & 15.82 \\
K60     & 16.72 & 16.20 & 15.95 \\
K61     & 14.11 & 13.92 & 13.81 \\
K62     & 17.46 & 17.39 & 17.14 \\
K63     & 17.17 & 17.74 & 17.24 \\
K64     & 16.86 & 16.45 & 16.32 \\
K65     & 14.47 & 13.88 & 13.79 \\
K66     & 14.73 & 14.36 & 14.13 \\
K68     & 13.67 & 13.44 & 13.29 \\
K69     & 14.92 & 14.46 & 14.35 \\
C3      & 14.04 & 13.41 & 13.11 \\
C4      & 13.15 & 12.71 & 12.55 \\
C5      & 16.98 & 16.62 & 16.64 \\
C6      & 15.71 & 15.08 & 14.82 \\
C7      & 16.26 & 15.48 & 15.27 \\
\enddata
\label{tab:table_candir}
\end{deluxetable}

\end{document}